\documentclass[sigconf,nonacm]{acmart}

\usepackage{url}
\usepackage{graphicx}
\usepackage{amsmath}
\usepackage{amsfonts}
\usepackage{paralist}
\usepackage{booktabs}
\usepackage{multirow}
\usepackage{multicol}

\usepackage{weiwMacro}
\usepackage{weiwMLmacros}
\usepackage{weiwAlgorithm}
\usepackage[tight,footnotesize]{subfigure}    
\usepackage{array}
\usepackage{cases}
\usepackage{color}
\usepackage{colortbl}
\definecolor{shadecolor}{gray}{0.85}
\newtheorem{lemma}{Lemma}
\newtheorem{problem}{Problem}

\newcommand{\etal}{\textit{et al.}\xspace}

\newcommand{\tmax}{t_{\max}}

\newcommand{\selnetrp}{$\textsf{SelNet}$\xspace}
\newcommand{\selnet}{$\textsf{SelNet}_{\text{-ct}}$\xspace}
\newcommand{\selnetnotau}{$\textsf{SelNet}_{\text{-ad-ct}}$\xspace}

\newcommand{\rs}{\textsf{RS}\xspace}
\newcommand{\is}{\textsf{IS}\xspace}
\newcommand{\qra}{\textsf{QR-1}\xspace}
\newcommand{\qrb}{\textsf{QR-2}\xspace}

\newcommand{\xgb}{\textsf{XGBoost}\xspace}
\newcommand{\minmax}{\textsf{MinMaxNet}\xspace}

\newcommand{\rmi}{\textsf{RMI}\xspace}
\newcommand{\dnn}{\textsf{DNN}\xspace}
\newcommand{\moe}{\textsf{MoE}\xspace}
\newcommand{\dln}{\textsf{DLN}\xspace}
\newcommand{\kde}{\textsf{KDE}\xspace}

\newcommand{\lightgbm}{\textsf{LightGBM}\xspace}

\newcommand{\umnn}{\textsf{UMNN}\xspace}
\newcommand{\cardnet}{\textsf{CardNet}\xspace}
\newcommand{\covertree}{\textsf{CoverTree}\xspace}
\newcommand{\faiss}{\textsf{Faiss}\xspace}
\newcommand{\exactsel}{\textsf{ExactSel}\xspace}
\newcommand{\truetime}{\textsf{TrueTime}\xspace}

\newcommand{\fasttext}{\textsf{fastText}\xspace}
\newcommand{\glove}{\textsf{GloVe}\xspace}
\newcommand{\face}{\textsf{MS-Celeb}\xspace}

\newcommand{\youtube}{\textsf{YouTube}\xspace}

\newcommand{\deep}{\textsf{DEEP}\xspace}
\newcommand{\sift}{\textsf{SIFT}\xspace}
\newcommand{\aminer}{\textsf{AMiner-Paper}\xspace}
\newcommand{\quora}{\textsf{Quora}\xspace}

\newcommand{\mse}{\textsf{MSE}\xspace}
\newcommand{\mre}{\textsf{MAPE}\xspace}
\newcommand{\mae}{\textsf{MAE}\xspace}
\newcommand{\mape}{\textsf{MAPE}\xspace}

\newcommand{\NNtau}{g^{(\tau)}}
\newcommand{\NNp}{g^{(p)}}

\newcommand{\LossEst}{J_{\text{est}}}

\newcommand{\confversion}[1]{}
\newcommand{\fullversion}[1]{#1}
\newcommand{\revise}[1]{#1}

\usepackage[all=normal, floats, bibnotes, wordspacing, charwidths, indent, lists]{savetrees}

\makeatletter%
\skip\footins 8.1pt plus 4pt minus 2pt % Space between last line of text top
\copyrightyear{2021}
\acmYear{2021}
\setcopyright{rightsretained}
\acmConference[SIGMOD '21] {Proceedings of the 2021 International Conference on Management of Data}{June 20--25, 2021}{Virtual Event, China}
\acmBooktitle{Proceedings of the 2021 International Conference on Management of Data (SIGMOD '21), June 20--25, 2021, Virtual Event, China}
\acmPrice{}
\acmISBN{978-1-4503-8343-1/21/06}
\acmDOI{10.1145/3448016.3452772}

\begin{document}

%\fancyhead{}

\title{Consistent and Flexible Selectivity Estimation for High-Dimensional Data} 

\newcommand{\sics}{$^{1}$}
\newcommand{\osaka}{$^{2}$}
\newcommand{\nagoya}{$^{3}$}
\newcommand{\dgut}{$^{4}$}
\newcommand{\unsw}{$^{5}$}
\newcommand{\unimelb}{$^{6}$}

\author{Yaoshu Wang\sics, Chuan Xiao\osaka$^,$\nagoya, Jianbin Qin\sics, Rui Mao\sics, Makoto Onizuka\osaka, Wei Wang\dgut$^,$\unsw, Rui Zhang\unimelb, and Yoshiharu Ishikawa\nagoya}
\affiliation{\sics{}Shenzhen Institute of Computing Sciences, Shenzhen University, \osaka{}Osaka University, \nagoya{}Nagoya University, 
\dgut{}Dongguan University of Technology, \unsw{}University of New South Wales, \unimelb{}\href{http://ruizhang.info}{www.ruizhang.info}}
\email{yaoshuw@sics.ac.cn, {chuanx, onizuka}@ist.osaka-u.ac.jp, {qinjianbin,mao}@szu.edu.cn, weiw@cse.unsw.edu.au, rui.zhang@ieee.org, ishikawa@i.nagoya-u.ac.jp}

%% weiw
\thanks{W.~Wang, R.~Mao and J.~Qin are the joint corresponding authors.}

%\thanks{\Letter~ are corresponding authors}

\begin{abstract}
% big picture and motivation
% problem specification
% key limitations in the literature
% main idea and innovation of this paper
% key findings and contribution of the paper

  % Selectivity estimation for high dimensional data is defined as follow.
  % Given a collection $\mathcal{D}$ of high dimensional objects
  % ($\mathbf{r} \in \mathbb{R}^d, \mathbf{r} \in \mathcal{D}$), a query
  % $\mathbf{x} \in \mathbb{R}^d$, a scalar threshold $\tau$, and a
  % distance function
  % $dist(\cdot, \cdot)$, estimate the selectivity of $\xx$, such as
  % $|\{\mathbf{r} | \mathbf{r} \in \mathcal{D}, dist(\mathbf{x},
  % \mathbf{r}) \leq \tau\}|$.
  Selectivity estimation aims at estimating the number of database objects that
  satisfy a selection criterion.
  % In this paper, we consider the case where
  % objects are high-dimensional vectors, and the selection criterion is 
  Answering this problem accurately and efficiently is essential to many 
  applications, such as density estimation, outlier detection, query 
  optimization, and data integration. %
  The estimation problem is especially challenging for large-scale
  high-dimensional data due to the curse of dimensionality, the large variance 
  of selectivity across different queries, and the need to
  make the estimator consistent (i.e., the selectivity is non-decreasing in 
  the threshold). 
  %range of the estimate that can be large.
  %especially when $|\mathcal{D}|$ is
  %large.
  We propose a new deep learning-based model that learns a \emph{query-dependent 
  piecewise linear function} as selectivity estimator, which is flexible to fit 
  the selectivity curve of any distance function and query object, while 
  guaranteeing that the output is non-decreasing in the threshold. 
  To improve the accuracy for large datasets, we propose to partition the dataset 
  into multiple disjoint subsets and build a local model on each of them. 
  % with a set of control points for
  % threshold to fit the curve of ground truth. We observe that control points
  % should be dependent on specific $\xx$. Based on this, we design a novel
  % network architecture that fully learns the generation of the best control
  % points in threshold and their predictions with implementation of neural
  % network.
  We perform experiments on real datasets and show that the proposed model 
  consistently outperforms state-of-the-art models in accuracy in an efficient 
  way and is useful for real applications. 
  % over two datasets and
  % three error metrics.
%Furthermore, considering the large
 %value of $|\mathcal{D}|$, we design a partition-based estimation
 %model to improve the performance. 
  % Empirical studies on two
  % real-world datasets verify that our method outperforms
  % state-of-the-art models and is more robust with
  % aspects of three
  % evaluation metrics.  
\end{abstract}

%%% Local Variables:
%%% mode: latex
%%% TeX-master: "paper"
%%% End:




%\settopmatter{printfolios=false}

\maketitle

\section{Introduction}

%%!! introduce rising trend of using AI in other areas, esp. in DB (Learned
%%index & SageDB)

%% emphasize on traditional approach's limitation.
In this paper, we consider the following selectivity estimation problem 
for high-dimensional data: 
given a query object $\xx$, a distance function $dist(\cdot, \cdot)$, and a 
distance threshold $t$, estimate the number of objects $\oo$s in a database 
that satisfy $dist(\xx, \oo) \leq t$. %% 
\revise{This problem is also known as local density estimation~\cite{wu2018local} 
or spherical range counting~\cite{DBLP:conf/soda/AryaMM05} in theoretical computer 
science. It is an essential procedure in density estimation in 
statistics~\cite{whang1994dynamic} and density-based outlier 
detection~\cite{DBLP:conf/sigmod/BreunigKNS00} in data mining. For example, for text 
analysis, one may want to determine the popularity of a topic; for e-commerce, an 
analyst may want to find out if a user/item is an outlier; for clustering, the 
algorithm may converge faster if we start with seeds in denser regions. In the 
database area, accurate estimation helps to find an optimal query execution plan in 
databases dealing with high-dimensional data~\cite{DBLP:conf/sigmod/HeimelKM15}. 
Hands-off entity matching systems~\cite{DBLP:conf/sigmod/DasCDNKDARP17} 
extract paths from random forests and take each path  -- a conjunction of similarity 
predicates over multiple attributes (e.g., ``$\textsf{EU}(\text{name}) 
\leq 0.25$ \textsf{AND} $\textsf{EU}(\text{affiliations}) \leq 0.4$ \textsf{AND} 
$\textsf{EU}(\text{research interests}) \leq 0.45$'', where $\textsf{EU}()$ measures 
the Euclidean distance between word embeddings) -- as a blocking rule, and efficient 
blocking can be achieved if we find a good query execution 
plan~\cite{DBLP:conf/sigmod/WangXQ0SWO20}. 
% Other applications include estimating the number of candidates and hence overall running 
% time to reach a service level agreement in an end-to-end image retrieval system, where 
% images are converted to embeddings~\cite{cao2017hashnet} so that a set of candidates can 
% be quickly obtained for image-level verification by sophisticated models (e.g., CNN). 
% In addition, many modern recommender systems resort to latent representations (embeddings) 
% of users and/or items~\cite{DBLP:journals/corr/Kula15,DBLP:conf/ijcai/YingZZLXXX018}. To 
% make recommendation, a selection query is invoked to obtain a set of candidates to be further 
% ranked by sophisticated models. Estimating the number of candidates helps to choose a proper 
% ranker to improve the quality of recommendation.
In addition, many text or image retrieval systems resort to distributed representations. 
Given a query, a similarity selection is often invoked to obtain a set of candidates to be 
further verified by sophisticated models. Estimating the number of candidates helps to 
estimate the overall query processing time an end-to-end system to create a service level 
agreement.}

% In data exploration, selectivity
% estimation measures the data distribution, skewness, and
% densities~\fixme{citation?}.

% Recently, rising trend of using neural networks in the database area has
% achieved significant performance, such as Recursive Model
% Index~\cite{kraska2018case} and \textsf{SageDB}~\cite{kraska2019sagedb}. It is
% necessary and promising to involve neural network into selectivity estimation
% problem, especially for high dimensional data.

Selectivity estimation for large-scale high-dimensional data is still an
open problem due to the following factors:
\begin{inparaenum}[(1)]
  \item \emph{Large variance of selectivity}. The selectivity varies across 
  queries and may differ by several orders of magnitude. A good estimator is 
  supposed to predict accurately for both small and large selectivity values. 
  \item \emph{Curse of dimensionality}. Many methods that work well on 
  low-dimensional data, such as histograms~\cite{ioannidis2003history}, are 
  intractable when we seek an optimal solution, and they significantly lose 
  accuracy with the increase of dimensionality.
  \item \emph{Consistency requirement}. When the query object is fixed, selectivity 
  is non-decreasing in the threshold. Hence users may want the estimated selectivity 
  to be non-decreasing and interpretable in applications such as density estimation. 
  This requirement rules out many existing methods. 
%   If strictly applied, this will rule out many methods. For example, isotonic or lattice
%   regressions need to be used.
\end{inparaenum}

% such as the curse of dimensionality. Existing
% methods fall into the following categories:
% \begin{inparaenum}[(i)]
% \item Sampling-based methods~\cite{wu2018local}.
% \item Kernel density estimation (KDE) based methods~\cite{DBLP:conf/edbt/MattigFBS18}.
% \item Machine learning-based methods~\cite{garcia2009lattice}.
% \end{inparaenum}

% There are several challenges for this problem:
%
% \begin{inparaenum}[(i)]
% \item High-dimensionality. 
% \item Wide range of output values, range from $[0, n]$, where $n$ is the number
%   of objects in the database. 
% \item Consistency. The estimator needs to be monotonic in the threshold
%   parameter. Let $\est{f}(q, t)$ be the estimated value. For distance functions,
%   this requires $\est{f}(q, t') \geq \est{f}(q, t)$ if and only if $t' \geq t$. 
% \end{inparaenum}

%% table of strength and weakness of existing methods wrt the above points? 

%% ours
To address the above challenges, we propose a novel deep regression method 
that guarantees consistency. 
% and provides a highly flexible and 
% accurate fit to the ground truth values. %
We holistically approximate the selectivity curve using a \emph{query-dependent 
piecewise linear function} consisting of control points that are learned from training 
data. This function family is \emph{flexible} in the sense that it can closely 
approximate the selectivity curve of any distance function and any input query 
object; 
%allows us to fit many functions closely,
e.g., using more control points for the part of the curve where selectivity 
changes rapidly. Together with a robust loss function, we are able to alleviate 
the impact of large variance across different queries. %
%and the optional partitioning optimization,
%we are able to alleviate the impact of a large range of the selectivity.
%
To handle high dimensionality, we incorporate an autoencoder that learns the 
latent representation of the query object with respect to the data distribution. 
The query object and its latent representation are fed to a query-dependent 
control point model, enhancing the fit to the selectivity curve of the query 
object. 
To ensure consistency, we achieve the monotonicity of estimated selectivity 
by converting the problem to 
%key control parameters by reducing it to 
a standard neural network prediction task, rather than imposing additional 
limitations such as restricting weights to be non-negative~\cite{daniels2010monotone} 
or limiting to multi-linear functions~\cite{fard2016fast}. %
To improve the accuracy on large-scale datasets, we propose a 
partition-based method to divide the database into disjoint subsets 
and learn a local model on each of them. 
Since update may exists in the database, we employ incremental learning to 
cope with this issue. 

%
% we learn the best control points for each
% query, and this idea further boosted the accuracy of our method. %
%
% 

We perform experiments on six real datasets. The results show that our method 
outperforms various state-of-the-art models. Compared to the best existing 
model~\cite{DBLP:conf/sigmod/WangXQ0SWO20}, the improvement of accuracy is up to 
5 times in mean squared error and consistent across datasets, distance 
functions, and error metrics. The experiments also demonstrate that our method is 
competitive in estimation speed, robust against update in the database, and 
useful in estimating overall query processing time in a semantic search application.

\section{Related Work}
\label{sec:related-work}

\paragraph{Traditional Estimation Models}
Selectivity estimation has been extensively studied in database systems, where
prevalent approaches are based on 
sampling~\cite{DBLP:conf/icde/WuAA02,wu2016sampling}, 
histograms~\cite{ioannidis2003history}, or 
sketches~\cite{DBLP:journals/ftdb/CormodeGHJ12}. However, few of them are
applicable to high-dimensional data due to data sparsity or the curse of 
dimensionality. For cosine similarity, Wu \etal~\cite{wu2018local} 
proposed to use locality-sensitive hashing (LSH) as a means of importance 
sampling to tackle data sparsity. 
%attain an improved unbiased estimation of selectivity. 
%This approach is heavily tied to a handful of 
%distance or similarity functions that have known LSH functions. %
%(e.g., there is no known LSH function for edit distance),
% In addition, the sample complexity is still inversely proportional to the
% selectivity, meaning that its estimate quality tends to be bad when only few
% objects satisfy the selection criterion. 
% \cite{wu2018local} adopted importance sampling strategy and multiple locality
% sensitive hashing to generate an unbiased estimator by constructing multiple
% buckets.
% the inverse proportion of the final selectivity, and are very loose. This leads
% to high estimation cost and underestimation issue of small selectivity values.
% Reasonably, Sampling-based approaches often combine with Locality sensitive
% hashing (\lsh) to reduce the complexity.
% \cite{spring2017new} designed a \lsh
% sampler to estimate the partition functions of log-linear models.
%The
%estimator was capable to handle high dimensional data with Cosine
%distance.
% Although \cite{wu2018local} reduces the complexity theoretically, it
% is still ineffective to handle underestimation problem.
%Sampling-based approaches often suffer from high estimation
%cost and underestimation issue of small cardinality result.
%For
%example, the bound of samples in~\cite{wu2018local} are inverse
%proportion to the cardinality value, which lead to inefficiency when
%cardinality is small.
Kernel density estimation (\kde)~\cite{DBLP:conf/sigmod/HeimelKM15,DBLP:conf/edbt/MattigFBS18} 
has been proposed to handle selectivity estimation in metric space. 
%It can be deemed as a smoothed version of sampling with learnable parameters. 
%(typically the kernel function). 
Mattig \etal~\cite{DBLP:conf/edbt/MattigFBS18} proposed to alleviat the curse of 
dimensionality by focusing on the distribution in metric space. However, 
strong assumptions are usually imposed on the kernel function (e.g., only 
diagonal covariance matrix for Gaussian kernels), and one kernel function 
may be inadequate to model complex distributions in high-dimensional data.

% Sampling-based approaches aim to maintain enough samples such that the
% selectivity in the database is estimated from these samples with theoretical
% guarantee. Sampling strategies determine the final performance.
% Sampling~\cite{DBLP:conf/sigmod/WuAA01}\fixme{xx} are basic sampling methods that
% adaptively control the sample sizes. Reservoir sampling is a random data
% selection method that scales up the estimation according to the proportion of
% samples in the dataset.

\paragraph{Regression Models without Consistency Guarantee}
Selectivity estimation can be formalized as a regression problem with
query object and threshold as input features, if the consistent requirement 
is not enforced. %
% Other deep models, e.g.,
% recursive model index~\cite{kraska2018case}, and mixture of Expert
% model~\cite{shazeer2017outrageously}, can also be adopted for
% regression with good performance due to their novel network
% structure.
\revise{A representative approach is quantized regression~\cite{DBLP:conf/icdm/Anagnostopoulos15a,DBLP:journals/tkdd/Anagnostopoulos17}.}
% quantize the query space to discover prototypes of query patterns and 
% estimate the selectivity by localized regression.
%Non-deep learning-based regression models (e.g, support vector regression) 
%, such as support vector regression and random forest, 
%However, they do not perform well for our task due to the high dimensionality. 
%Deep
%regression models~\cite{lathuilire2018comprehensive} are widely used
%in the area of computer vision, such
%as human pose estimation, age estimation, and so forth. However, they
%focus on extracting valuable features from images or videos to improve
%performance.
Recent trend uses deep regression models. %
Vanilla deep regression~\cite{lathuiliere2018comprehensive,toshev2014deeppose,sun2013deep}
learns good representations of input patterns. 
%and uses a fully connected regression layer to predict. 
The mixture of expert model (\moe)~\cite{shazeer2017outrageously} has a sparsely-gated 
mixture-of-experts layer that assigns data to proper experts (models) 
%and each expert is deterministically
%and separately learnt. \moe has
which lead to better generalization. The recursive model index 
(\rmi)~\cite{kraska2018case} is a regression model that 
can be used to replace the B-tree index in relational 
databases. Deep regression has also been used to predict  
%~\cite{DBLP:journals/corr/abs-1808-03196,DBLP:journals/pvldb/MarcusNMZAKPT19,DBLP:conf/sigmod/MarcusP18},
selectivities (cardinalities)~\cite{DBLP:conf/cidr/KipfKRLBK19,DBLP:journals/pvldb/SunL19} 
%and query performance~\cite{DBLP:conf/icde/AkdereCRUZ12} 
in relational databases, amid a set of recent advances in learning 
methods for this task~\cite{DBLP:journals/corr/abs-1905-06425,DBLP:journals/pvldb/YangLKWDCAHKS19,DBLP:journals/pvldb/WalenzSRY19,DBLP:conf/sigmod/HasanTAK020,DBLP:conf/sigmod/ParkZM20}. 
\revise{They target SQL queries where each predicate involves one attribute. 
\cite{DBLP:conf/sigmod/HasanTAK020,DBLP:journals/pvldb/YangLKWDCAHKS19} employ 
autoregressive models. \cite{DBLP:conf/sigmod/ParkZM20} only deals with low 
dimensionality. \cite{DBLP:conf/cidr/KipfKRLBK19,DBLP:journals/pvldb/SunL19,DBLP:journals/corr/abs-1905-06425} 
become a deep neural network if we regard a vector as an attribute.}
% When used to replace the traditional B-tree index, its
% main task is to
% predict the location of the first entry whose key value is at least a given
% query key value. Due to its hierarchical model structure, \rmi is suitable for
% solving regression problem with a wide output
% range. % which is one feature of our problem
%The above regression methods do not guarantee consistency in terms of monotonicity.
% monotonicity can be enforced at the splitting time. 

% Robust regression~\cite{huber1964robust,lathuiliere2018deepgum} focuses on
% handling outliers caused by $L_2$ loss function, and achieves robust regression.

\paragraph{Models with Consistency Guarantee}
%
% Although
%There are only a few regression models that guarantee consistency. %
Gradient boosting trees (e.g., \xgb~\cite{chen2016xgboost} 
and \lightgbm~\cite{DBLP:conf/icbsp/WangZZ17}) support monotonic regression. 
% Isotonic regression~\cite{han2017isotonic,spouge2003least} fits a free-form
% monotonic curve to a set of training data. As it is non-parametric, it is not
% clear how to adapt it to our selectivity estimation problem where the output is
% only partially monotonic in its input. %
% \minmax~\cite{daniels2010monotone} is a monotonic neural network that designs a
% min-max layer to ensure monotonicity.
% % Although it is proved to satisfy the
% % universal approximation theorem, it is not flexible and suitable enough for
% % regression with large output space.
% Training is typically slow as only a small part of the network gets non-zero
% gradient during back propagation, and its accuracy is not satisfactory~\cite{you2017deep}. %
Lattice regression~\cite{garcia2009lattice,fard2016fast,gupta2016monotonic,you2017deep} 
uses a multi-linearly interpolated lookup table for regression. %
% while ensuring monotonicity on all or part of the input features.
% interpolated function based on the parameters associated with the vertices of a
% lattice.
%
By enforcing constraints on its parameter values, it can guarantee
monotonicity. To accommodate high dimensional inputs, Fard \etal~\cite{fard2016fast}
proposed to build an ensemble of lattice using subsets of input features. 
%The choice of the ensemble is determined either heuristically or randomly. 
Deep lattice network (\dln)~\cite{you2017deep} was proposed to interlace non-linear 
calibration layers and ensemble of lattice layers. % in order to be more flexible and deeper. 
Recently, lattice regression has also been used to learn a spatial index~\cite{DBLP:conf/sigmod/Li0ZY020}. 
\umnn~\cite{unconstraintmono} is an autoregressive flow model which adopts 
Clenshaw-Curtis quadrature to achieve 
%the integral of uni-variable and guarantee 
monotonicity. Other monotonic models include isotonic regression~\cite{han2017isotonic,spouge2003least} 
and \minmax~\cite{daniels2010monotone}. 
% Isotonic regression~\cite{han2017isotonic,spouge2003least} fits a free-form monotonic 
% curve to a set of training data, but as it is non-parametric, it is not clear how to 
% adapt it to our selectivity estimation problem. \minmax~\cite{daniels2010monotone} is 
% a monotonic neural network having a min-max layer to ensure monotonicity, but the 
% training is typically slow as only a small part of the network gets non-zero gradient 
% during back propagation, and its accuracy is not satisfactory~\cite{you2017deep}. 
\cardnet~\cite{DBLP:conf/sigmod/WangXQ0SWO20} is a recently proposed method for monotonic 
selectivity estimation of similarity selection query for various data types. 
It maps original data to binary vectors and the threshold to an integer $\tau$, and 
then predicts the selectivity for distance $[0, 1, \ldots, \tau]$ respectively with 
$(\tau + 1)$ encoder-decoder models. When applying to high-dimensional data, its has 
the following drawbacks: the mapping from the input threshold to $\tau$ is not injective, 
i.e., multiple thresholds may be mapped to the same $\tau$ and the same selectivity is 
always output for them; the overall accuracy is significantly affected if one of the 
$(\tau + 1)$ decoders is not accurate for some query.

% We collectively name them \lattice in the rest of the paper. 

% our model learns a rich class of functions (continuous piecewise linear
% function) 
% \lattice
% fails to explore how to construct interpolated look-up tables (or calibrated
% layer) according to other non-monotonic features. In other words, \lattice fails
% to consider partitioning threshold according to different record
% $\mathbf{x}$.\fixme{not well arguement.}
% The idea of prediction with control points has been adopted in \lattice
% model~\cite{you2017deep,gupta2016monotonic,fard2016fast}, a suite of latest
% monotonic models.

% Different from them, our model fully
% learns a good set of control points of monotonic feature, i.e., threshold,
% and achieves more flexiblity. Furthermore, we adopt a delicately designed
% function to learn interpolation values of control points instead of directly
% treating them as parameters. Our model has better generalization
% and is capable to handle complex selectivity estimation, and we will
% discuss it in detail in Section~\ref{sec:estimator}.
\section{Preliminaries}
\label{sec:prelim}

%In this section, we define the problem and some notations, and introduce a brief
%review of related work.

%\subsection{Definitions and Notations}

%Similarity functions can be easily transferred to
%corresponding distance ones, e.g., $sim_{cos}(\cdot, \cdot) = 1 -
%dist_{cos}(\cdot, \cdot)$, so we only discuss distance functions in
%the rest of the paper for simplicity.

\begin{problem}[Selectivity Estimation for High-Dimensional Data]
  Given a database of $d$-dimensional vectors $\mathcal{D} = \set{\oo_i}_{i=1}^n, \oo_i \in \mathbb{R}^d$, 
  a distance function $dist(\cdot, \cdot)$, a scalar threshold $t$, and a query 
  object $\xx \in \mathbb{R}^d$, estimate the selectivity in the database, i.e., 
  $| \set{\oo \mid dist(\xx, \oo) \leq t, \oo \in \mathcal{D}}|$.
\end{problem}
While we assume $d$ is a distance function, it is easy to extend it to consider
$d$ as a similarity function by changing $\leq$ to $\geq$ in the above definition. 
In the rest of the paper, to describe our method, we focus on the case when $d$ 
is a distance function. 
%, while we evaluate Euclidean distance and cosine similarity in our experiments. 
\revise{In addition, the query object does not have to be in the database, and we 
do not make any assumption on the distance function, meaning that the function 
does not have to be metric.}
% For ease of exposition, we assume

% We consider a distance function
% $d: \mathbb{R}^m \times \mathbb{R}^m \to \mathbb{R}$, such that a scalar
% distance value measures the distance of two $m$-dimensional vectors. $L_p$
% ($p \geq 1$) distance and Cosine distance are commonly used in the machine
% learning community.

We can view the selectivity (i.e., the ground truth label) $y$ of a query object $\xx$ 
and a threshold $t$ as generated by a function $y = f(\xx, t, \mathcal{D})$. 
We call $f$ the \textbf{value function}.
Our goal is to estimate $f(\xx, t, \mathcal{D})$ using another function $\est{f}(\xx, t, \mathcal{D})$.

One unique requirement of our problem is that the estimator $\est{f}$ needs to
be \emph{consistent}: $\est{f}$ is consistent if and only if it is \emph{non-decreasing} 
in the threshold $t$ for every query object $\xx$; i.e., 
$\forall \xx$, $\est{f}(\xx, t',  \mathcal{D}) \geq \est{f}(\xx, t, \mathcal{D})$ iff. $t' \geq t$.

%%% Local Variables:
%%% mode: latex
%%% TeX-master: "paper"
%%% End:

% !TeX root = body/ARN.tex

\section{Observations and Ideas}
\label{sec:idea}

When $|\mathcal{D}|$ is large, it is hard to estimate $f$ directly. One of the
main challenges is that $f$ may be non-smooth with respect to the input variables. 
In the worst case, we have: 
%$f$ is not smooth with respect to the input variables.
%% put them as Lemma? 
\begin{itemize}
\item For any vector $\Delta\xx$, there exists a database $\mathcal{D}$ of $n$ objects and a query
  $(\xx, t)$ such that $f(\xx, t, \mathcal{D}) = 0$ and $f(\xx + \Delta\xx, t, \mathcal{D}) = n$.
\item For any $\epsilon > 0$, there exists a database $\mathcal{D}$ of $n$ objects and a query
  $(\xx, t)$ such that $f(\xx, t, \mathcal{D}) = 0$ and $f(\xx, t+\epsilon, \mathcal{D}) = n$.
\end{itemize}
%% need logic and support
This means any model that directly approximates $f$ is hard.

\revise{Our idea to mitigate this issue is: instead of estimating one function $f$, we estimate 
multiple functions such that each function's output range is a small fraction of the selectivity 
$y$. For example, suppose $y = y_1 + y_2$ and $t = t_1 + t_2$. If $y_1$ and $y_2$ are
approximately linear in $[0, t_1]$ and $(t_1, t_2]$, respectively, but with different 
slopes, then we can use two linear models for the two threshold ranges. We may also exploit 
this idea and divide $y$ with disjoint subsets of $\mathcal{D}$.} 
Hence we adopt the following two partitioning schemes. 

\paragraph{Threshold Partitioning}
Assume the maximum threshold we support is $t_{\max}$. We consider dividing it with an increasing 
sequence of $(L + 2)$ values: 
$[\tau_0, \tau_1, \dots, \tau_{L + 1}]$ such that $\tau_{i} < \tau_{j}$ if $i < j$, $\tau_0 = 0$, and 
$\tau_{L + 1} = t_{\max} + \epsilon$, where $\epsilon$ is a small positive quantity~\footnote{$\epsilon$ 
is used to cover the corner case of $t = t_{\max}$ in Eq.~\eqref{eq:interpolation}.}. 
% we let $\Delta \tau_j \definedas \tau_{j} - \tau_{j-1}$, and $g_{j}(\xx, \Delta \tau_j, \mathcal{D}) 
% \definedas f(\xx, \tau_{j}, \mathcal{D}) - f(\xx, \tau_{j-1}, \mathcal{D})$. 
% Then we have $\est{f}(\xx, t, \mathcal{D})= \sum_{i=1}^{N}{g_i(\xx, \Delta \tau_i, \mathcal{D})}$.
Let $g_i(\xx, t)$ be an interpolant function for interval $[\tau_{i-1}, \tau_i)$. Then we have 
\begin{align}
  \label{eq:interpolation}
  \est{f}(\xx, t, \mathcal{D}) = \sum_{i=1}^{L+1} \Indicator{t \in [\tau_{i-1}, \tau_i)} \cdot g_i(\xx, t), 
\end{align}
where $\Indicator{}$ denotes the indicator function. 
    
\paragraph{Data Partitioning}
We partition the database $\mathcal{D}$ into $K$ disjoint parts $\mathcal{D}_1, \dots, \mathcal{D}_K$, 
and let $f_i$ denote the value function defined on the $i$-th part. 
Then we have $\est{f}(\xx, t, \mathcal{D}) = \sum_{i=1}^K{\est{f_i}(\xx, t, \mathcal{D}_i)}$.

\section{Selectivity Estimator}\label{sec:estimator}

%In this section, we introduce our model for selectivity estimation. 

\subsection{Threshold Partitioning}
%We first consider the value function $f(\xx, t)$ at a fixed $\xx$.
Our idea is to approximate $f$ using a regression model $\est{f}(\xx, t, \mathcal{D}; \Theta)$. 
% $\est{f}$, which is a 
% quadratic spline, denoted by $\est{f}(\xx, t, \mathcal{D}; \Theta)$. The 
% quadratic spline has $(L + 2)$ control (interpolation) points, where $L$ is 
% the number of internal control points and $2$ refers to the two ends. %
% The parameters in $\est{f}(\xx, t, \mathcal{D}; \Theta)$ are dependent on $\xx$, 
% i.e., they are query-dependent. 
%%!! allows us to 'jump' in estimate
Recall the sequence $[\tau_0, \tau_1, \dots, \tau_{L + 1}]$ in Section~\ref{sec:idea}. 
% Let $p_i$ denote the estimated selectivity for a query object $\xx$ and a threshold $\tau_i$. 
% We assume the maximum threshold we support is $t_{\max}$. To cover all $t \in [0, \tmax]$, we 
% consider $(L + 2)$ values in the sequence where $L$ is the number of interval 
% values and $2$ refers to the two ends, i.e., $\tau_0 = 0$ and $\tau_{L+1} = \tmax$. 
% parameterized by 
% $\Theta \definedas \set{(\tau_i, p_i)}_{i=0}^{L+1}$, where $\tau_i$ and $p_i$ values 
% are dependent on $\xx$, i.e., they are query-dependent. 
We consider the family of continuous piecewise linear function to implement the interpolation 
$g_i(\xx, t)$, $i \in [0, L + 1]$. 
A piecewise linear function is a continuous function of $(L + 1)$ pieces, each being a linear 
function defined on $[\tau_{i-1}, \tau_i)$. 
The $\tau_i$ values are called \emph{control points}. Given a query object $\xx$, let $p_i$ 
denote the estimated selectivity for a threshold $\tau_i$. For the $g_i$ function in 
Eq.~\eqref{eq:interpolation}, we have 
% Moreover, we demand the quadratic spline be smooth; i.e., every  
% two adjacent pieces defined on $[\tau_{i-1}, \tau_i)$ and $[\tau_{i}, \tau_{i + 1})$ share 
% a common derivative value at $\tau$. 
% We choose such design for interpolation for the following two reasons: 
% \begin{inparaenum}
%   \item Universal approximation...
%   \item Smoothness is good for training...
% \end{inparaenum}
\begin{align}
%   & \est{f}(\xx, t, \mathcal{D}; \Theta) = \sum_{i=1}^{L+1} \Indicator{t \in [\tau_{i-1}, \tau_i)} \cdot
%     g_i(\xx, t), \label{eq:plf-predict}\\
%   & \text{where } 
  g_i(\xx, t) = p_{i-1} + \frac{t - \tau_{i-1}}{\tau_i - \tau_{i-1}} \cdot (p_i - p_{i-1}). %\nonumber.
  \label{eq:plf-predict}
%% actually, can be written as (1-z)P_i + zP_i
\end{align}
%We assume the maximum threshold we support is $t_{max}$. To cover all 
%$t \in [0, \tmax]$, we require $\tau_0 = 0$ and $\tau_{L+1} = \tmax$. 
Hence the regression model is parameterized by 
$\Theta \definedas \set{(\tau_i, p_i)}_{i=0}^{L+1}$. Note that $\tau_i$ and $p_i$ values are 
dependent on $\xx$; i.e., the piecewise linear function is query-dependent. 

Using the above design for $\Theta$ has the following property to guarantee the 
\confversion{consistency~\footnote{Proof is provided in Appendix A of the extended version~\cite{DBLP:journals/corr/abs-2005-09908}.}.}
\fullversion{consistency~\footnote{Proof is provided in Appendix~\ref{sec:proof}.}.}
\begin{lemma}
  \label{lm:mono}
  Given a database $\mathcal{D}$ and a query object $\xx$, if 
  $p_i \geq p_{i-1}$ for $\forall i \in [1, L + 1]$, 
  then $\est{f}(\xx, t, \mathcal{D}; \Theta)$ is non-decreasing in $t$. 
\end{lemma}
% In addition, it has the following property for flexibility.
% \begin{lemma}
%   \label{lm:flex}
%   Given a database $\mathcal{D}$ and a query object $\xx$, for any $\epsilon > 0$, 
%   there exists as an approximate realization $\est{f}(\xx, t, \mathcal{D}; \Theta)$ 
%   such that $|f(\xx, t, \mathcal{D}) - \est{f}(\xx, t, \mathcal{D}; \Theta)| < \epsilon$ 
%   for all $t$.
% \end{lemma}

% \begin{proof}
% We discuss the approximation power of piecewise linear function such that 
% $|f - \est{f}| = K|\Delta| \cdot |\nabla \est{f}|$, 
% where $K$ is a constant value, and $|\Delta| = \max_{i=0}^L\{\tau_{i+1} - \tau_i\}$. 
% Because $\est{f}$ is the piecewise linear function, $|\nabla \est{f}|$ is bounded
% by $\max_{i=0}^L\{\frac{p_{i+1} - p_i}{\tau_{i+1} - \tau_i}\}$. With more control points, i.e., larger $L$,
% and optimal $\{\tau_i\}_{i=1}^L$ and $\{p_i\}_{i=0}^{L+1}$, $K|\Delta| \cdot |\nabla \est{f}| < \epsilon$. 
% Second, $\{\tau_i\}_{i=1}^L$
% and $\{p_i\}_{i=0}^{L+1}$ are estimated by neural networks, and $K|\Delta||\nabla \est{f}|$ 
% is bounded by $K (|\Delta| + \epsilon_1 )\cdot (|\nabla \est{f}| + \epsilon_2)$. Due to the universal
% approximation of neural networks, there exist $\epsilon_1$ and $\epsilon_2$ such that
% $K(|\Delta| + \epsilon_1) (|\nabla \est{f}| + \epsilon_2) < \epsilon$.
% \end{proof}

%Our model does not rely on the triangle inequality. So it can support non-metric functions. 
Another salient property of our model is that it is flexible in the sense that it can arbitrarily 
well approximate the selectivity curve. Piecewise linear functions have been well explored to fit 
one-dimensional curves~\cite{prunty1983curve}. With a sufficient number of control points, one can 
find an optimal piecewise linear function to fit any one-dimensional curve. The idea is that a 
small range of input is highly likely to be linear with the output. When $\xx$ and $\mathcal{D}$ 
are fixed, the selectivity only depends on $t$, and thus the value function can be treated as a 
one-dimensional curve. To distinguish different $\xx$, we will design a deep learning approach to 
learn good control points and corresponding selectivities. As such, our model not only inherits 
the good performance of piecewise linear function but also handles different query objects.

\paragraph{Estimation Loss} 
In the regression model, the $L$ $\tau_i$ values and the $(L + 2)$ $p_i$ values are the 
parameters to be learned. 
We use the expected loss between $f$ and $\est{f}$: 
\begin{align}
  \LossEst(\est{f}) = \sum_{((\xx, t), y) \in
  \mathcal{T}_{\text{train}}}\ell(f(\xx, t, \mathcal{D}), \est{f}(\xx, t, \mathcal{D})), %= %
  % \int_{u=0}^{t_{\max}(\xx)} \ell(f(\xx, u), \est{f}(\xx, u)) P(u \mid \xx) \dd{u}, 
  \label{eq:loss-est}
\end{align}
where $\mathcal{T}_{\text{train}}$ denotes the set of training data, and 
$\ell(y, \est{y})$ is a loss function between the true selectivity $y$ 
and the estimated value $\est{y}$ of a query $(\xx, t)$. 
We choose the Huber loss~\cite{huber1964robust} 
%on the ratio of the estimate. Let $r \definedas \frac{\est{y}}{y}$ 
applied to the logarithmic values of $y$ and $\est{y}$. To prevent numeric
errors, we also pad the input by a small positive quantity $\epsilon$. Let 
$r \definedas \ln(y + \epsilon) - \ln(\est{y} + \epsilon)$. Then 
% where
% $\varepsilon$ is a small constant value to guarantee $r$ is valid.
\begin{align*}
  \ell(y, \est{y}) =
  \begin{cases}
    \frac{r^2}{2} & \text{, if } \abs{r} \leq \delta; \\
    \delta (\abs{r}  - \frac{\delta}{2}) & \text{, otherwise.}
  \end{cases}
\end{align*}
$\delta$ is set to $1.345$, the standard recommended value~\cite{Fox02robustregression}. %
The reason for designing such a loss function is that the selectivity may 
differ by several orders of magnitude for different queries. 
%differ may span several orders of magnitude when the database is large. 
%Typically we are
%interested in selectivity in the range of $[1, %\frac{n}{100}]$. 
If we use the $\ell_2$ loss, it encourages the model to fit large selectivities 
well, and if we use $\ell_1$ loss, it pays more attention to small selectivities. 
To achieve robust prediction, we reduce the value range by logarithm and the Huber 
loss.

\subsection{Learning Piecewise Linear Function}
% We need to construct a model to learn the parameters of the model, i.e., $L$ 
% $\tau_i$ values and $(L+2)$ $p_i$ values, all in a query-dependent way. 
We choose a deep neural network to learn the piecewise linear function. It has the 
following advantages: 
\begin{inparaenum} [(1)]
  \item Deep learning is able to capture the complex patterns in control points and 
  corresponding selectivities for accurate estimation of different queries. 
  \item Deep learning generalizes well on queries that are not covered by training 
  data. 
  \item The training data for our problem can be unlimitedly acquired by running 
  a selection algorithm on the database, and this favors deep learning which 
  often requires large training sets. 
\end{inparaenum}

In our model, $\tau_i$ and $p_i$ values are generated separately for the input query 
object. We also require non-negative increments between consecutive parameters to 
ensure they are non-decreasing. %
In the following, we explain the learning of $\tau_i$s and $p_i$s, followed by the 
overall neural network architecture.
 
% \paragraph{Generating $\Theta_G$}

% Our idea is to directly learn the parameters of truncated Guassian,
% i.e., 
% $\Theta_G \definedas \{\mu, \sigma, k, a, b\}$ as follows.
% \begin{align*}
% [\mu, \sigma, k, a', b'] = \relu(\NNg(\xx))
% \end{align*}
% where $\NNg(\xx)$ is a function to output $\Theta_G$ with query point
% $\xx$ as input. $\relu$ is used due to the nonnegativity of these
% values. Here we do not directly learn the interval, i.e., $[a, b]$ of
% truncated Guassian but offsets, such that $a = \mu - a'$ and $b = \mu
% + b'$. The reason is that we expect to learn a truncated Guassian
% distribution such that its mode can be included and the breaking ranges of
% selectivity curves (red circles in Figure~\ref{fig:eg-obs}) might be approximated. 

\paragraph{Control Points ($\tau_i$s)}
We learn the increments between $\tau_i$s. %Specifically, 
%We model $\tau_i$ for a given $\xx$ as:
% \begin{align*}
%   & \tau_{i}(\xx) = \sum_{j=0}^{i-1} \BDelta_{\tau}(\xx)[j] \qqtext{,  where}\\
%   & \BDelta_{\tau}(\mathbf{x}) = \Softmax(\NNtau(\xx)) \cdot t_{max}
% \end{align*}
\begin{align}
  & \tau_{i}(\xx) = \sum_{j=0}^{i-1} \BDelta_{\tau}(\xx)[j], \\%\qqtext{,  where}\\
  & \text{where } \BDelta_{\tau}(\mathbf{x}) = \text{Norm}_{l_2}(\NNtau(\xx)) \cdot t_{max}.
  \label{eq:learn-tau}
\end{align}
$\text{Norm}_{l_2}$ is a normalized squared function defined as
\revise{
\begin{align*}
  \text{Norm}_{l_2}(\tt) = [\frac{t_1^2 + \frac{\epsilon}{L}}{\tt^{\rm T}\tt +
  \epsilon}, \dots, \frac{t_L^2 + \frac{\epsilon}{L}}{\tt^{\rm T}\tt +
  \epsilon}],
\end{align*}
}
where $\epsilon$ is a small positive quantity to avoid dividing by 
zero, %$d$ is the dimensionality of $\tt$, 
%set as $10^{-6}$. 
and $t_i$ denotes the value of the $i$-th dimension of $\tt$. 
The model takes $\xx$ as input and outputs $L$ distinct thresholds in $(0, \tmax)$. 
$\NNtau$ is implemented by a neural network. Then we have a vector 
$\Btau = [0; \tau_1; \tau_2; \dots; \tau_L ; t_{max}]$.
%$\t{\mqty[0 & \tau_1 & \ldots & \tau_L & \tmax]}$. 

One may consider using $\Softmax(\tt)$, which is widely
used for multi-classification and (self-)attention. 
We choose $\text{Norm}_{l_2}(\tt)$ rather than $\Softmax(\tt)$ for 
the following reasons: 
\begin{inparaenum} [(1)]
  \item Due to the exponential function in $\Softmax(\tt)$,
  a small change of $\tt$ might lead to large variations of the
  output.
  \item $\Softmax$ aims to highlight the important part 
  rather than partitioning $\tt$, while our goal is to rationally 
  partition the range $[0, \tau_{max}]$ into several intervals such 
  that the piecewise linear function can fit well. 
\end{inparaenum}
% For example,
% multi-classification learns one important class and attention more
% focuses on finding more correlated alignments. In the scenario of
% generating $\tau_i$, 
%Thus, $\text{Norm}_{l_2}$ is more suitable than $\Softmax$.

\paragraph{Selectivities at Control Points ($p_i$s)}
We learn $(L + 2)$ $p_i$ values in a similar fashion to control points, 
using another neural network to implement $\NNp$.
\begin{align}
  & p_{i}(\xx) = \sum_{j=0}^{i} \BDelta_{p}(\xx)[j], \\ %\qqtext{,  where}\\
  & \text{where } \BDelta_{p}(\mathbf{x}) = \relu(\NNp(\xx)). 
  \label{eq:learn-p}
\end{align}
Then we have a vector $\pp = [p_0; p_1; \dots; p_{L+1}]$.
Here, we learn $(L + 1)$ increments ($p_i - p_{i - 1}$)
instead of directly learning $(L + 2)$ $p_i$s. Thereby, 
we do not have to enforce a constraint $p_{i-1} \leq p_i$
for $i \in [1, L + 1]$ in the learning process, and thus the learned 
model can better fit the selectivity curve. 

\begin{figure}[t]
  \centering
  \includegraphics[width=\linewidth]{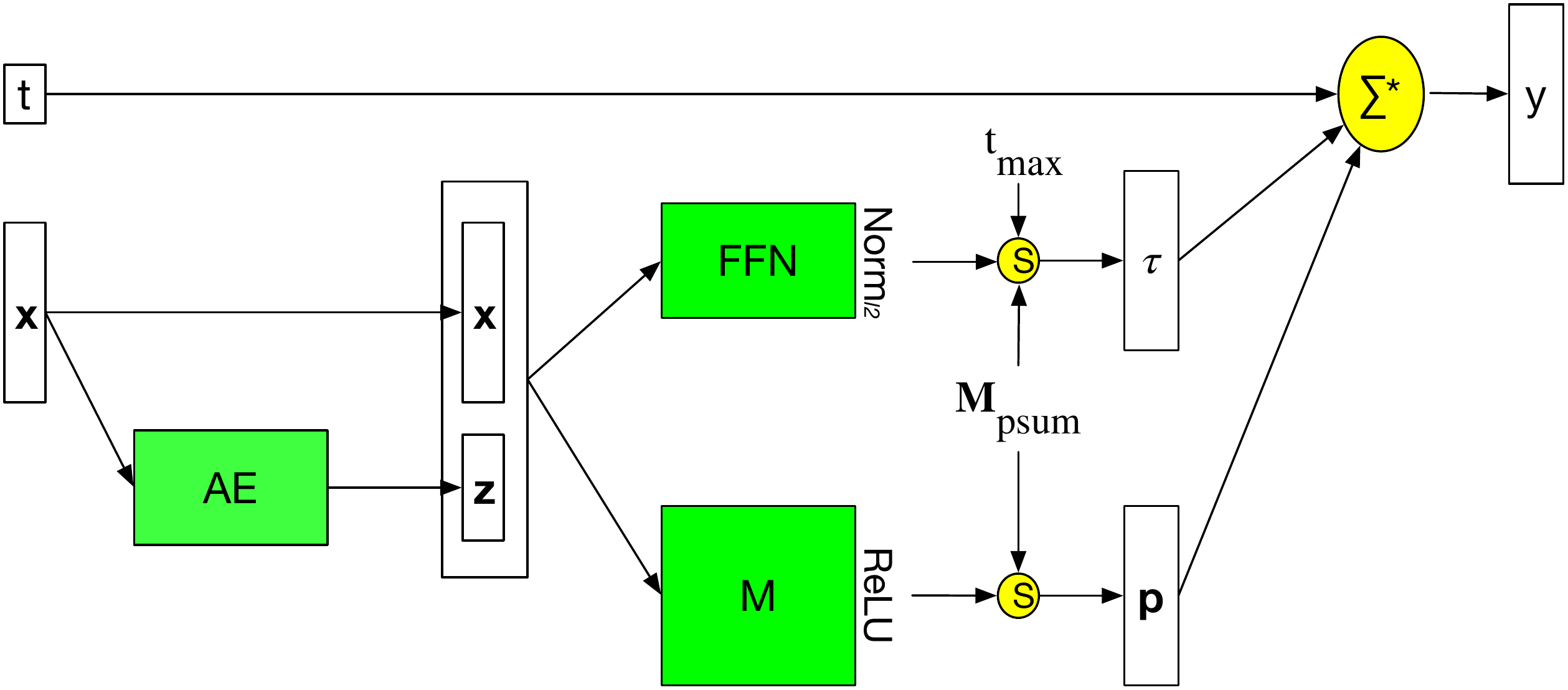}
  \caption{Network architecture.}
  \label{fig:network}
\end{figure}

\paragraph{Network Architecture}
Figure~\ref{fig:network} shows our network architecture.
% As mentioned above, two functions
% $\NNtau$ and $\NNp$ need to be learned by using neural
% networks. 

% model $\NNtau$ and $\NNp$

%\vspace{-2mm}

The input $\xx$ is first transformed to %$\smqty[\xx \\ \zz]$, 
$\zz$, a latent representation obtained by an autoencoder (AE). 
%on the entire database. %
% Justify VAE
The use of the AE encourages the model to exploit latent data and query
distributions in learning the piecewise linear function, and this helps the model 
generalize to query objects outside the training data. 
%, such that the AE is able to learn the hidden differences between $\xx$ and $\xx + \epsilon$.
%Here notice that we aims to dig the hidden information of the database $\mathcal{D}$, 
To learn the latent distributions of $\mathcal{D}$, we pretrain the AE on all 
the objects of $\mathcal{D}$, and then continue to train the AE with the queries 
in the training data. 
%Similar design has been used in previous models for binary data and integer
%thresholds~\cite{our-integer-est}. 
% continue
% The encoder part of the model transform the input $(\xx, \tau)$ to
% $\t{\mqty[\xx ; \zz_{\xx}]}$\fixme{...}, which is then feed into two different
Due to the use of AE, the final loss function is a linear combination of the 
estimation loss (Eq.~\eqref{eq:loss-est}) and the loss of the AE for the 
training data (denoted by $J_{\text{AE}}$):
\begin{align}
  J(\est{f}) = \LossEst(\est{f}) + \lambda \cdot J_{\text{AE}}. \label{eq:final-loss}
\end{align}
%%!! Note: AE has been pretrained. 

$\xx$ is concatenated with $\zz$, i.e., $[\xx; \zz]$. Then $[\xx; \zz]$ is fed into two 
independent neural networks: a feed-forward network ($\text{FFN}$) and a model $M$ 
(introduced later). 
%We labelled the last layer activation functions as well. 
Two multiplications, denoted by $S$ operators in Figure~\ref{fig:network}, 
are needed to separately convert the output of $\text{FFN}$ and the output of model 
$M$ to the $\Btau$ and $\pp$ vectors, respectively. They use a scalar $\tmax$ 
and a matrix $\MM_{\text{psum}}$ which, once multiplied on the right to a vector, 
perform prefix sum operation on the vector.
\begin{align*}
  \MM_{\text{psum}} =
  \begin{bmatrix}
    1 & 0 & \ldots & 0 \\
    1 & 1 & \ldots & 0 \\
    \vdots & \vdots & \vdots & \vdots\\
    1 & 1 & \ldots & 1
  \end{bmatrix}.
\end{align*}

The output of these networks, together with the threshold $t$, are fed into the 
operator $\sum^*$ in Figure~\ref{fig:network}, which is implemented by Eqs.~\eqref{eq:plf-predict}, 
~\eqref{eq:learn-tau}, and~\eqref{eq:learn-p}, to compute the output of the piecewise 
linear function, i.e., the estimated selectivity.
% The $\sum^*$ operator computes the estimates based on
% Equation~\eqref{eq:plf-predict}. Furthermore, partial outputs, i.e.,
% $\{\mu, \sigma, k, a, b\}$ of $M$ 
% are used as parameters of the cumulative distribution of truncated
% Guassian and its estimate is outputed. The final estimation is the sum
% of the estimate of piecewise linear function and truncated Guassian. 

\paragraph{Model $M$}
To achieve better performance, we learn $\pp$ using an 
encoder-decoder model. In the encoder, an FFN is used to generate $(L + 2)$ 
embeddings: 
\begin{align}\label{eq:modelm}
  [\hh_0; \hh_1; \dots; \hh_{L + 1}] = \text{FFN}([\xx; \zz]), 
\end{align}
where $\hh_i$s are high-dimensional representations. Here, we adopt 
$(L + 2)$ embeddings, i.e., $\hh_0, \dots, \hh_{L + 1}$, to represent the
latent information of $\pp$.
In the decoder, we adopt $(L + 2)$ linear transformations with the \text{ReLU} 
activation function: 
\begin{align*}
  k_i = \relu(\ww_i^{\rm T}\hh_i + b_i). 
\end{align*}
Then we have $\pp = [k_0, k_0 + k_1, \dots, \sum_{i=0}^{L+1}{k_{i}}]$. 

% $\mathrm{MLP}_2$ is designed in the fashion of multi-tasking
% learning. Learning $L + 1$ incremental predictions is
% considered as learning $L + 1$ tasks. To achieve that, we adopt the
% strategy of sharing a deep model, which generates $L + 1$ task
% embeddings. After that, $L + 1$ decoders with corresponding task
% embeddings as inputs
% are discriminatively trained
% to generate $L + 1$ incremental predictions. Here, we implement the deep model as
% a feedforward neural network, and decoder as linear transformation
% with \textsf{ReLU} activation function.  

% \paragraph{Deterministic Prediction}

% We note that the mapping from $\xx$ to $\zz_{\xx}$ contains the randomness
% injected by VAE. We keep such randomness in the training so that the latent
% representation $\zz_{\xx}$ can generalize well. However, to ensure the
% consistency of the learned function, at prediction time, we remove such
% randomness by using the expected latent representation. Similar approximation is
% also used in~XXX\fixme{cite the paper}

% \comment{
% \paragraph{Discussions}

% We can show that the model learned in our architecture is guaranteed to be
% consistent, i.e.,
% $\est{f}(\xx, \tau) \leq \est{f}(\xx, \tau + \epsilon), \forall \xx, \tau, \epsilon \geq
% 0$. This is mainly due to the we learned non-negative increments on $\tau_i$ and
% $p_i$ due to the $\Softmax$ and $\relu$ functions.
% %\fixme{also due to the determinisitinc use of VAE in prediction.}
% }

%\subsection{Training Data Generation}

\subsection{Data Partitioning}

% The input space should be related with the original data
% distribution. Some outliers $\mathbf{x}$ are far away from any data in
% $\mathcal{D}$ or located in sparse regions, then they have relatively small
% cardinality values in high probability. Furthermore, for $\mathbf{x}$
% that is located in dense regions of $\mathcal{D}$, its cardinality
% value is large.

% One unique challenge in the selectivity estimation is that the value function
% may have very large partial derivatives in both input parameters. That is, a
% small change in $\xx$ or $t$ may lead to a large change in the selectivity values.
% This makes it hard to learn. %
% A small disturbance $\epsilon$ in $\xx$ might lead to large difference of
% selectivities, such that $|f(\xx, \tau) - f(\xx + \epsilon, \tau)|$ is large.
To improve the accuracy of estimation on large-scale datasets, we divide the 
database into multiple disjoint subsets $\mathcal{D}_1, \dots, \mathcal{D}_K$ 
with approximately the same size, and build a local model on each of them. Let 
$\est{f}_i$ denote each local model. Then the global model for selectivity 
estimation is $\est{f} = \sum_i \est{f}_i$.

%\paragraph{Special Designs}

We have considered several design choices and propose the following configuration 
that achieves the best empirical performance:
\begin{inparaenum}[(1)]
  \item Partitioning is obtained by a cover tree-based strategy. 
  %and %\footnote{Note that this can %be done on-the-fly} 
  \item We adopt the structure in Figure~\ref{fig:network} so that all local models 
  share the same input representation $[\xx; \zz]$, %$\smqty[\xx\\ \zz]$, 
  but each has its own neural networks to learn the control points.
\end{inparaenum}

\begin{figure} [t]
  \centering
    \includegraphics[width=.6\linewidth]{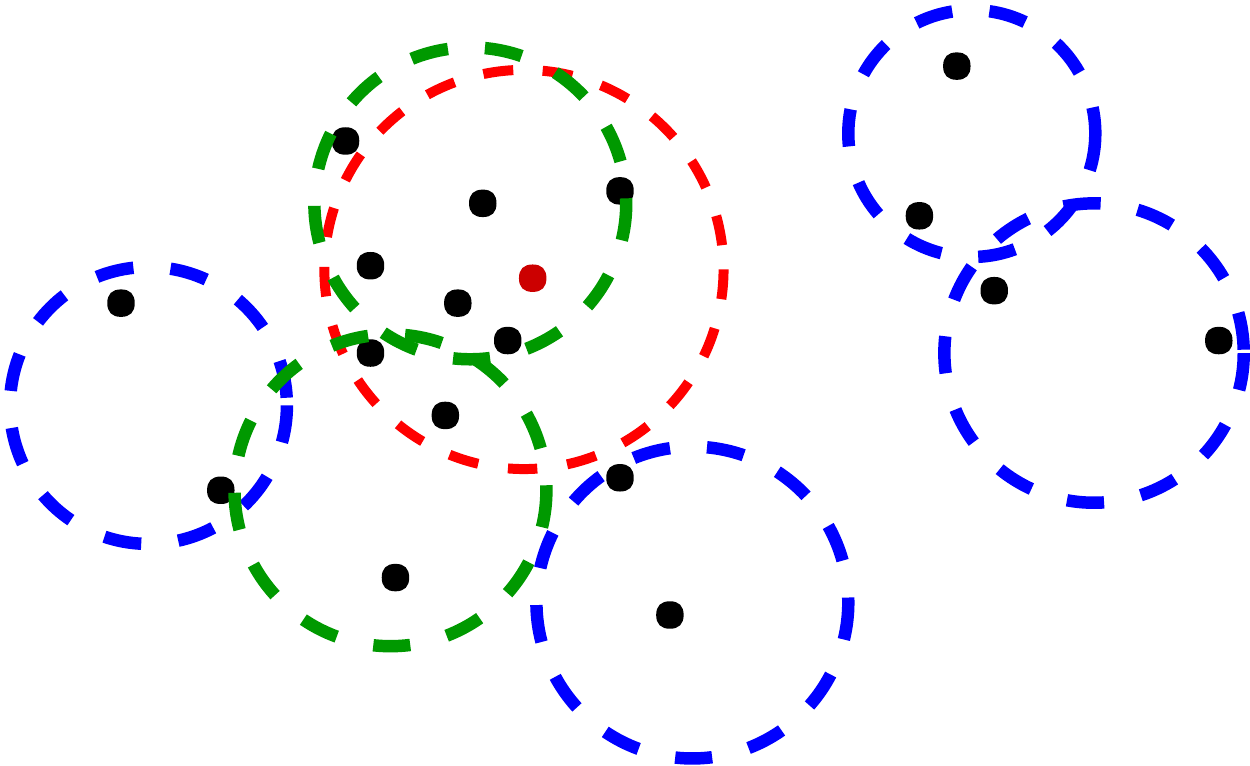}
  \caption{Data partitioning by cover tree.}
  \label{fig:covertree}
\end{figure}

% \begin{align}\label{eq:prediction}
% \widehat{y} = \sum_{i=1}^{K}{\mathbf{c}[i] \cdot
%   \mathcal{F}^i(\mathbf{x}, \tau)}
% \end{align}

\paragraph{Partitioning Method}
We utilize a cover tree~\cite{fastercovertree} to partition $\mathcal{D}$ 
into several parts. A partition ratio $r$ is predefined such that the cover tree will 
not expand its nodes if the number of inside objects is smaller than $r|\mathcal{D}|$. 
Given a query $(\xx, t)$, the valid region is the circles that intersect the circle 
with $\xx$ as center and $t$ as radius. For example, in Figure~\ref{fig:covertree}, 
$\xx$ (the red point) and $t$ form the red circle, and data are partitioned into 6 
regions. The valid region of $(\xx, t)$ is the green circles that intersect the red 
circle. Albeit imposing constraints, cover tree might still generate too many ball 
regions, i.e., leaf nodes, which lead to large number of parameters of the model and 
the difficulty of training. Reducing the number of ball regions is necessary. To 
remedy this, we adopt a merging strategy as follows. First, we still partition 
$\mathcal{D}$ into $K'$ regions using cover tree. Then we cluster these regions into 
$K$ ($K' \leq K$) clusters $\mathcal{D}_1, \dots, \mathcal{D}_K$ by the following 
greedy strategy: The $K'$ regions are sorted in decreasing order of the number of 
inside objects. We begin with $K$ empty clusters. Then we scan each region and 
assign it to the cluster with the smallest size. The regions that belong to the same 
cluster are merged to one region. We consider an indicator 
$f_c: (\xx, t) \to \set{0, 1}^{K}$ such that $f_c(\xx, t)[i] = 1$ 
if and only if the query $(\xx, t)$ intersects cluster $\mathcal{D}_i$, and employ 
it in our model: 
\begin{align*}
  \est{f}(\xx, t, \mathcal{D}) = \sum_{i=0}^K{f_c(\xx, t)[i] \cdot \est{f}_i(\xx, t, \mathcal{D}_i)}.
\end{align*}
% In the applications of selectivity estimation, metric distance functions (e.g., 
% Euclidean distance) that support triangle inequality are often utilized.
Since cover trees deal with metric spaces, for non-metric functions (e.g, cosine 
similarity), if possible, we equivalently convert it to a metric (e.g, Euclidean 
distance, as 
$\cos(\uu, \mathbf{v}) = 1 - \frac{\norm{\uu, \mathbf{v}}^2}{2}$ for unit vectors 
$\uu$ and $\mathbf{v}$). Then the cover tree partitioning still works. For those 
that cannot be equivalently converted to a metric, we adopt random partitioning 
and modify $f_c$ as $f_c: (\xx, t) \to \set{\mathbf{1}}^K$. 

\paragraph{Training Procedure}

We have several choices on how to train the models from multiple partitions. %
The default is directly training the global model $\est{f}$, with the
advantage that no extra work is needed. The other choice is to train each local
model independently, using the selectivity computed on the local partition as
training label. We propose yet another choice: we pretrain the local models
for $T$ epochs, and then train 
them jointly. In the joint training stage, we use the following loss function:
% \begin{align*}
%   J_{\text{joint}} = J(\est{f}^{(*)}) +  \beta \cdot \sum_i J(\est{f}^{(i)}).
% \end{align*}
\begin{align*}
  J_{\text{joint}} = \LossEst(\est{f}) +  \beta \cdot \sum_i \LossEst(\est{f}_i) + \lambda \cdot J_{\text{AE}}.
\end{align*}
The indicators $f_c(\cdot, \cdot)$s of all $(\xx, t)$ are precomputed before training. 

\subsection{Dealing with Data Updates}\label{sec-update}
When the database $\mathcal{D}$ is updated with insertion or 
deletion, we first check whether our model $\est{f}(\xx, t, \mathcal{D})$ is
necessary to update. In other words, when minor updates occur and
$\est{f}(\xx, t, \mathcal{D})$ is still accurate enough, we ignore them. To check
the accuracy of $\est{f}(\xx, t, \mathcal{D})$, we update the labels of all validation
data, and re-test the mean absolute error ($\text{MAE}$) of $\est{f}(\xx, t, \mathcal{D})$. If the
difference between the original $\text{MAE}$ and the new one is no larger
than a predefined threshold $\delta_U$, we do not update our model. 
Otherwise, 
% When $\est{f}(\xx, t, \mathcal{D})$ cannot make well estimates (larger differences
% than $\delta_U$), 
we adopt an incremental learning approach as follows. 
First, we update the labels in the training and the 
validation data to reflect the update in the database. Second, we
continue training our model with the updated training data until
the validation error ($\text{MAE}$) does not increase in 3
consecutive epochs. Here the training does not start from
scratch but from the current model. We incrementally train
our model with all the training data to prevent catastrophic
forgetting. 
% Meanwhile, when one record is updated via insertion,
% deletion or modification, we store it in an update set $\Delta
% \mathcal{D}$. If $|\Delta \mathcal{D}| \geq c$, where $c$ is a
% predefined cardinality value, we sample $k$ records from $\Delta
% \mathcal{D}$, calculate their labels and add them into training
% data. Here notice that new added data should not be included in the
% validation and testing datasets.

%%% Local Variables:
%%% mode: latex
%%% TeX-master: "paper"
%%% End:

\section{Discussions}
\label{sec:discussion}
% We compare our model with two representative existing monotonic models: 
% lattice regression and Clenshaw-Curtis quadrature. 

\subsection{Model Complexity Analysis}
We assume an FFN has hidden layers $\mathbf{a}_1, \dots, \mathbf{a}_n$. 
%As discussed in \cite{DBLP:conf/sigmod/WangXQ0SWO20}
The complexity of an FFN with input $\mathbf{x}$ and output $\mathbf{y}$ is 
$|\text{FFN}(\mathbf{x}, \mathbf{y})| = |\mathbf{x}| \cdot |\mathbf{a}_1| + 
\sum_{i=1}^{n-1}{|\mathbf{a}_i|\cdot |\mathbf{a}_{i+1}|} + 
|\mathbf{a}_n| \cdot |\mathbf{y}|$. 
% \begin{align*}
%     |\text{FNN}(\mathbf{x}, \mathbf{y})| = |\mathbf{x}| \cdot |\mathbf{a}_1|
%     + \sum_{i=1}^{n-1}{|\mathbf{a}_i|\cdot |\mathbf{a}_{i+1}|} +
%     |\mathbf{a}_n| \cdot |\mathbf{y}|.
% \end{align*}

Our model contains three components: AE, FFN, and $M$. 
The complexity of AE is $|\text{FFN}(\mathbf{x}, \mathbf{z})|$. The complexity of
FFN is $|\text{FFN}([\mathbf{x}; \mathbf{z}], \mathbf{t})|$, where $\mathbf{t}$ 
is the $L$-dimensional vector after $\text{Norm}_{l_2}$. Component $M$ consists of 
an FFN and $(L + 2)$ linear transformations. Its complexity 
is $|\text{FFN}([\mathbf{x}; \mathbf{z}], \mathbf{H})| + 
(L + 2) \cdot |\mathbf{h}_i| + (L + 2)$, where $\mathbf{H} = [\mathbf{h}_0; \dots; \mathbf{h}_{L+1}]$.
Thus, the final model complexity is
$|\text{FFN}(\mathbf{x}, \mathbf{z})| + |\text{FFN}([\mathbf{x}; \mathbf{z}], \mathbf{t})| + 
|\text{FFN}([\mathbf{x}; \mathbf{z}], \mathbf{H})| + (L + 2) \cdot |\mathbf{h}_i| + (L + 2)$.

% \subsection{Generality of Piece-wise Linear Functions}
% Piece-wise linear functions are well explored to fit a one-dimensional curve~\cite{prunty1983curve}. 
% With enough control points, we can find an optimal set of piece-wise linear functions to fit a 
% one-dimensional curve. The idea of piece-wise linear function is that a small range of input $X$ is 
% highly likely to be linear with output $Y$. We fully utilize its advantage. For each specific $\xx$, 

\subsection{Comparison with Other Models}
\label{sec:compare-lattice-regression}

\begin{figure} [t]
  \centering
  \subfigure[Simplified \dln{}]{
    \includegraphics[width=0.48\linewidth]{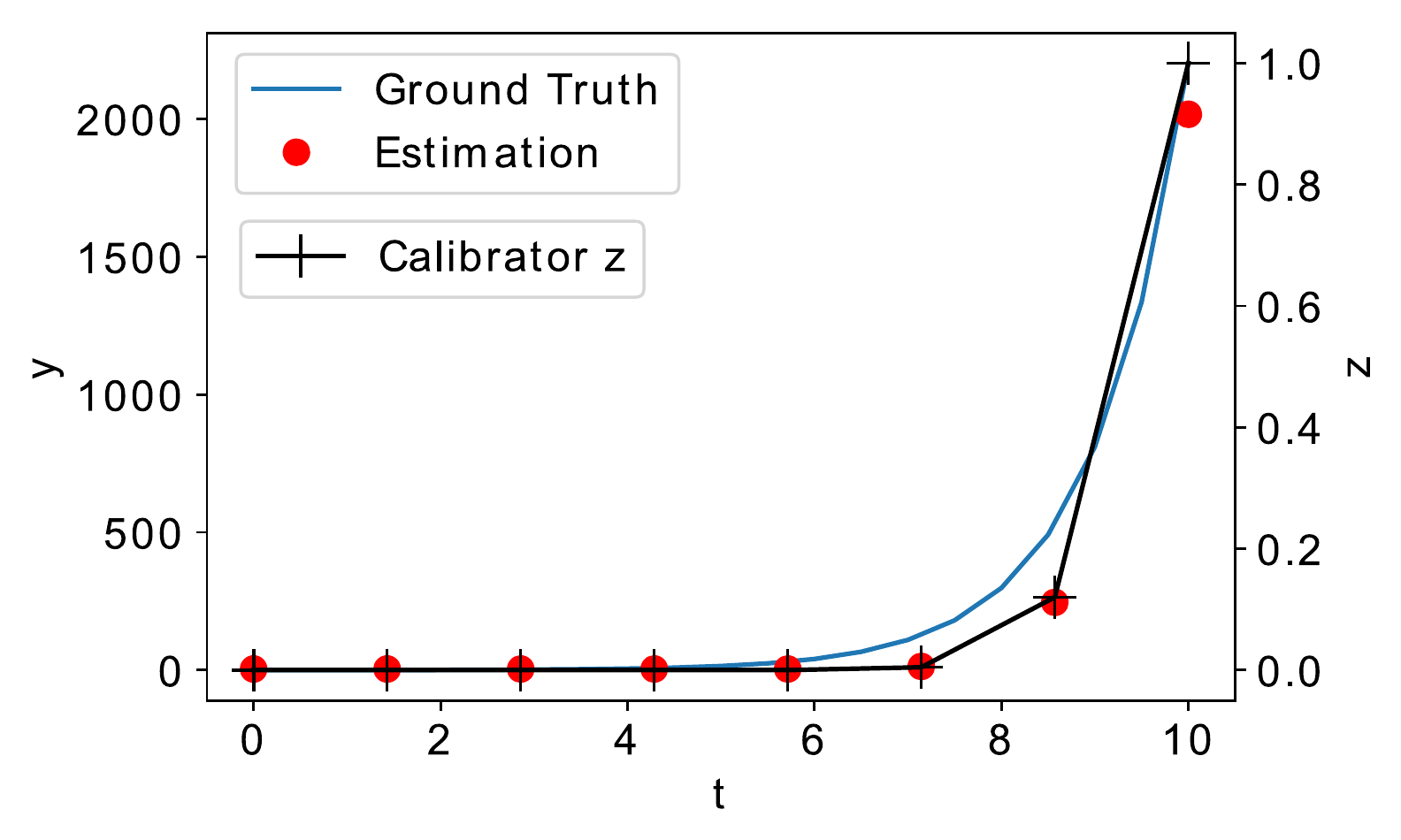}
    \label{fig:eg-lattice}
  }
  \subfigure[Our Model]{
    \includegraphics[width=0.45\linewidth]{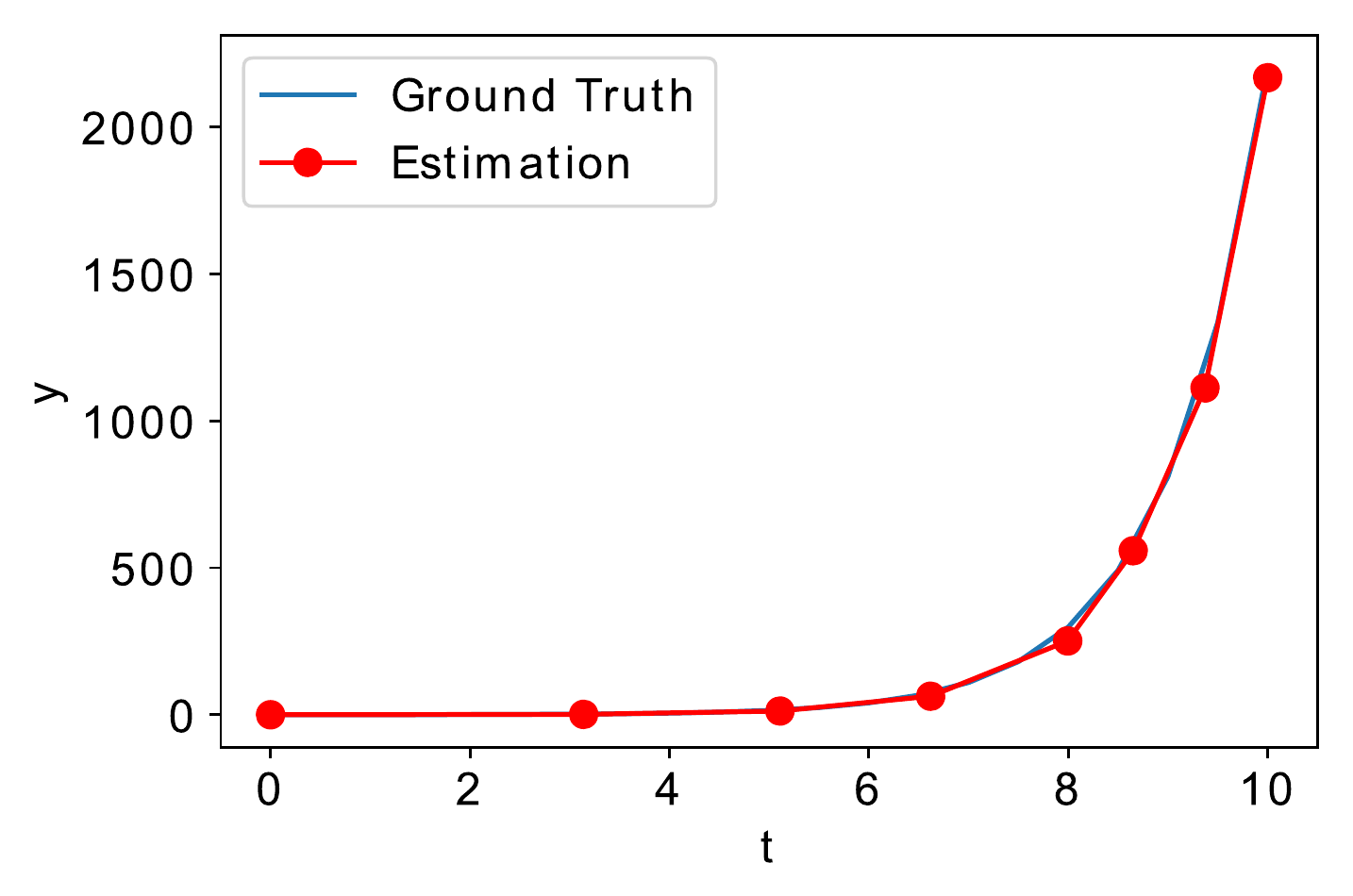}
    \label{fig:eg-selnet}
  }
  \caption{Comparison of simplified \dln{} and our model.} 
    %(both with 8 control points) to $y = f(t) = \frac{1}{10}\exp(t)$ for $t \in [0, 10]$.}
  \label{fig:eg-inter}
  %\vspace{-1em}
\end{figure}

\paragraph{Lattice Regression}
Lattice regression models~\cite{garcia2009lattice,fard2016fast,gupta2016monotonic,you2017deep} 
are the latest deep learning architectures for monotonic regression. We provide a comparison 
between ours and them applied to selectivity estimation. For the sake of an analytical 
comparison, we assume $\xx$ and $\mathcal{D}$ are fixed so the selectivity only depends on $t$, 
and consider a shallow version of \dln{}~\cite{you2017deep} with one layer of calibrator and one 
layer of a single lattice. 

With the above simplification, the \dln{} can be analytically represented
as: $\est{f}_{\text{DLN}}(t) = h(g(t; \ww); \theta_0, \theta_1)$, where 
$g: t \in [0, \tmax] \mapsto z \in [0, 1]$ and $h(z; \theta_0, \theta_1) = (1-z) \theta_{0} + z \theta_{1}$.
% \begin{align*}
%   & \est{f}_{\text{DLN}}(t) = h(g(t; \ww); \theta_0, \theta_1) \text{, where}\\
%   & \quad g: t \in [0, \tmax] \mapsto z \in [0, 1], \\
%   & \quad h(z; \theta_0, \theta_1) = (1-z) \theta_{0} + z \theta_{1}.
% \end{align*}
Hence it degenerates to fitting a linear
interpolation in a latent space. There is little learning for the
function $h$, as its two parameters $\theta_{0}$ and $\theta_{1}$ are
determined by the minimum and maximum selectivity values in the training data.
Thus, the workhorse of the model is to learn the non-linear mapping of $g$. 
%This has two inherent limitations:
The calibrator also uses piecewise linear functions with $L$ control points  
equivalent to our $(\tau_i, p_i)_{i=1}^L$. However, $\tau_i$s are equally spaced 
between $0$ and $\tmax$, and only $p_i$s are learnable. This design is not 
flexible for many value functions; e.g., if the function values change rapidly 
within a small interval, the calibrator will not adaptively allocate more control 
points to this area. We show this with 8 control points for both models to learn 
the function $y = f(t) = \frac{1}{10} \exp(t)$, $t \in [0, 10]$. The training 
data are 80 $(t_i, f(t_i))$ pairs where $t_i$s are uniformly sampled in $[0, 10]$. 
We plot both models' estimation curves and their learned control points in 
Figure~\ref{fig:eg-inter}. The $z$ values at the control points of \dln{} are 
shown on the right side of Figure~\ref{fig:eg-lattice}. We observe: 
\begin{inparaenum}[(1)]
  \item The calibrator virtually determines the estimation as $h()$ degenerates
  to a simple scaling.
  \item The calibrator's control points are evenly spaced in $t$, while our
  model learns to place more controls points in the ``interesting area'',
  i.e., where $y$ values change rapidly. 
  %This effect will be even more
  %pronounced if we consider the fact that the interesting area varies
  %depending on the query $\xx$.
  \item As a result, our model approximates the value function much better than
  \dln{}.
\end{inparaenum}

Further, for \dln, the non-linear mapping on $t$ is independent of $\xx$ (even 
though we do not model $\xx$ here). Even in the full-fledged \dln{} model, the 
calibration is performed on each input dimension independently. 
The full-fledged \dln{} model is too complex to analyze, so we only
study it in our empirical evaluation. Nonetheless, we believe that the above
inherent limitations still remain. Our empirical evaluation will also show that 
query-dependent fitting of the value function is critical in our problem. 
%, as it significantly improves the performance.
Apart from \dln, recent studies also employ lattice regression 
and/or piecewise linear functions for learned index~\cite{DBLP:conf/sigmod/Li0ZY020,DBLP:conf/sigmod/KipfMRSKK020}. 
Like \dln, their control points are also query independent, albeit not equally 
spaced. 

\fullversion{
%\subsection{Comparison with Clenshaw-Curtis Quadrature}
\paragraph{Clenshaw-Curtis Quadrature}
Clenshaw-Curtis quadrature~\cite{novelinkova2011comparison} is able to approximate the integral 
$\int_{0}^{\tau_{max}}{\est{g}(\xx, t, \mathcal{D})}\text{d}t$,
where $\est{g} = \frac{\partial \est{f}(\xx, t, \mathcal{D})}{\partial t}$ in our problem.
\umnn~\cite{unconstraintmono} is a recent work that adopts the idea
to solve the autoregressive flow problem, and uses a neural network
to model $\est{g}$. In \cite{novelinkova2011comparison}, the cosine 
transform of $\est{g}(cos \theta)$ is adopted and the discrete 
finite cosine transform is sampled at equidistant points $\theta = \frac{\pi s}{N}$, where 
$s = 1, \dots, N$, and $N$ is the number of sample points. Similar to \dln, it adopts
the same integral approximation for different queries and ignores that integral points 
should depend on $\xx$. In contrast, our method addresses this issue by using a 
query-dependent model, thereby delivering more flexibility. 
%Section~\ref{sec:exp-abalation} will
%give more detail explanations later.
}

\fullversion{
\paragraph{Query-Driven Quantized Regression} 
The main idea of query-driven quantized regression~\cite{DBLP:conf/icdm/Anagnostopoulos15a,DBLP:journals/tkdd/Anagnostopoulos17} 
is to quantize the query space and find prototypes (the closest 
one or multiple related ones) for the given query object. Then 
the output space is quantized by prototypes, and localized 
regressions are used to estimate the selectivity for 
corresponding prototypes. Like our model, they also employ a 
query-dependent design. The differences from ours are: 
\begin{inparaenum} [(1)]
  \item \cite{DBLP:conf/icdm/Anagnostopoulos15a,DBLP:journals/tkdd/Anagnostopoulos17} 
  divide the query space of $(\xx, t)$ while we divide the range 
  of threshold $t$ using $\xx$. 
  \item The number of prototypes is finite and often up to thousands 
  in \cite{DBLP:conf/icdm/Anagnostopoulos15a,DBLP:journals/tkdd/Anagnostopoulos17}, 
  while our model chooses the selectivity curve for the query object 
  via an FFN and model $M$ (Figure~\ref{fig:network}), which yield an 
  unlimited number of curves in $\mathbb{R}^{L+2}$. 
  \item We employ deep regression for higher accuracy. 
  \item We directly partition the database $\mathcal{D}$ and 
  train multiple deep models to deal with the subsets of 
  $\mathcal{D}$ that may differ in data distribution, while the data 
  subspace in \cite{DBLP:journals/tkdd/Anagnostopoulos17} is defined by 
  its query prototype. 
  %\item The methods in \cite{DBLP:conf/icdm/Anagnostopoulos15a,DBLP:journals/tkdd/Anagnostopoulos17} 
  %are not monotonic. 
\end{inparaenum}
% The main idea is to quantize the query space and selectivity space. Although 
% both of \selnetrp and them are query-driven, they mainly focus on finding the 
% closest {\em winning} prototype of the given query. Also, after constructing 
% multiple query prototypes $Q$, the output domain is quantized according
% to $Q$. Our model is much different from them as follows. First, we adopt
% the query-driven idea to partition the given threshold, such that different
% query point learns different threshold partitions of piecewise linear functions
% implemented by neural networks, which is not considered by \cite{DBLP:conf/icdm/Anagnostopoulos15a} and \cite{DBLP:journals/tkdd/Anagnostopoulos17}.
% Second, we partition the original dataset $\mathcal{D}$ instead of 
% quantizing the query and output space.
% We observe that the output domain is $[0, |\mathcal{D}|]$, so our main goal is to partition the output space by partitioning $\mathcal{D}$, while \cite{DBLP:conf/icdm/Anagnostopoulos15a} and \cite{DBLP:journals/tkdd/Anagnostopoulos17} mainly link the output values, i.e., labels, to corresponding query prototypes.

% Thus, \selnetrp uses the query-driven strategy to handle thresholds (which is one of the most important factors of selectivity estimation problem), and 
% uses the data-centric approach to handle large-scale datasets to increase
% the flexibility of models.
}
\section{Evaluations}
\label{sec:exp}
\confversion{Please see \cite{DBLP:journals/corr/abs-2005-09908} for detailed setup and additional experiments. 
%(source codes and datasets are also available at this GitHub repository)}
}
%The source code is available at \cite{source-code}.

%In this section, we report and analyze our experimental results. 
%In this section, we report our experimental results.

% investigate the performance of our model  to state-of-the-art methods. We mainly focus on
% whether our novel network architecture outperforms others by designing
% experiments to verify as follow.
% \begin{inparaenum}[(i)]
% \item the overall accuracy of our network architecture, 
% \item 
% the effect of function to learn several incremental predictions, i.e., $\NNp$,
% \item the improved threshold partition model $\NNtau$,
% \item random
% partition strategy \textsf{RP}.
% \end{inparaenum}

\begin{table} [t]
  \small%\scriptsize
  \caption{Statistics of datasets.}
  \label{tab:dataset}
  \centering
  \resizebox{\linewidth}{!}{
  \begin{tabular}[b]{|l|c|c|c|c|c|}
    \hline%
    Dataset & Source & Domain & \# Objects & Dimensionality & Distance \\
    \hline%
    \fasttext & \cite{URL:fasttext} & text & 1M & 300 & Euclidean \\
    \glove & \cite{URL:glove} & text & 1.9M & 300 & Euclidean \\
    \face & \cite{guo2016msceleb} & image & 2M & 128 & cosine \\
    \youtube & \cite{URL:youtube-faces} & video & 0.35M & 1770 & cosine \\
    \deep & \cite{URL:deep} & image & 100M & 96 & cosine \\
    \sift & \cite{URL:sift} & image & 200M & 128 & cosine \\
    \hline%
  \end{tabular}
  }
\end{table}

\subsection{Experimental Settings}

\paragraph{Datasets}%
\revise{We use six datasets. The statistics is given in Table~\ref{tab:dataset}. 
We preprocess \face by \textsf{faceNet}~\cite{schroff2015facenet} to obtain vectors. 
The other datasets have already been transformed to high-dimensional data. \glove, 
\youtube, \deep, and \sift were also used in previous work~\cite{DBLP:conf/sigmod/WangXQ0SWO20} 
or nearest neighbor 
search benchmarks~\cite{DBLP:journals/is/AumullerBF20,DBLP:journals/tkde/LiZSWLZL20}.}

We randomly sample 0.25M vectors from each dataset $\mathcal{D}$ as query objects. %
%%!! mention that thresholds needs to be different for different x
\confversion{
\revise{To generate thresholds, we sample a geometric sequence (by uniformly sampling from 
logarithmic values) of 40 selectivity values in the range $[1, 1\% \size{\mathcal{D}}]$. 
Then for each selectivity $y$, we calculate the minimum threshold that yields at least $y$ 
results.}
}
% \fullversion{
% We consider two settings to generate thresholds: 
% \begin{inparaenum} [(1)]
%   \item For the default setting,  
%   we sample a geometric sequence (by uniformly sampling from logarithmic values) of 40 
%   selectivity values in the range $[1, 1\% \size{\mathcal{D}}]$. Then for each selectivity 
%   $y$, we calculate the minimum threshold that yields at least $y$ results.
%   %; e.g., the cosine similarity for $1\% \size{\mathcal{D}}$ is around 0.5 on \fasttext. 
%   \item For the alternative setting, we generate cosine similarity thresholds in $[0.2, 1]$ 
%   using the beta distribution $\alpha = 3$ and $\beta = 2.5$, in order to 
%   simulate the case when people are more interested in thresholds in the middle. 
%   %The selectivity is up to around $10\% \size{\mathcal{D}}$ on \fasttext. 
% %   The selectivities of queries change rapidly within the high probability 
% %   region of the distribution, thus yielding larger variance than the default setting. 
% %   The selectivity is up to $10\% \size{\mathcal{D}}$, with a cosine similarity of around 0.2 
% %   on fasttext. 
% \end{inparaenum}
% }

The resulting query workload, denoted by $\mathcal{Q}$, was uniformly split in 8:1:1 (by 
query objects) into training, validation, and test sets. So none of the test query objects has been seen 
by the model during training or validation. Note that labels (i.e., true selectivities) are 
computed on $\mathcal{D}$, not $\mathcal{Q}$. 
\fullversion{For each training query object, we iterate through all the generated thresholds 
and add them to the training set. We randomly choose 3 generated thresholds for each 
validation or test query object. Due to the large number of training data, we randomly select 
training instances for each batch instead of continuously loading them, and the training 
procedure terminates when the mean squared error of the validation set does not increase in 5 
consecutive epochs.} 
For each setting, we tested on 5 sampled workloads to mitigate the effect of sampling error.

\paragraph{Methods} %
We compare the following %our model with six state-of-the-art models, covering a variety of
\confversion{approaches.}
\fullversion{approaches~\footnote{Please see Appendix~\ref{sec:exp-setup} for model settings.}.}
\begin{itemize} %[noitemsep,wide=0pt, leftmargin=\dimexpr\labelwidth + 2\labelsep\relax]
  \item \revise{\textbf{\rs} is a random sampling approach. For each query, we uniformly sample 
  $0.1\% \size{\mathcal{D}}$ objects for the first four datasets and $0.01\% \size{\mathcal{D}}$ 
  objects for \deep and \sift. Then we use \textsf{scipy.spatial.distance.cdist} to compute 
  the distances to the query objects in a batch manner.}
  \item \textbf{\is}~\cite{wu2018local} is an importance sampling approach using
  locality-sensitive hashing. %that substantially reduces the number of samples required. 
  It only works for cosine similarity due to the use of SimHash~\cite{charikar2002similarity}. %technique. 
  %We did not perform the Euclidean distance to cosine conversion as
  %its performance is not competitive. 
  We enforce monotonicity by using deterministic sampling w.r.t. the query 
  object.
  \item \textbf{\kde}~\cite{DBLP:conf/edbt/MattigFBS18} is based on adaptive
  kernel density estimation for metric distance functions. 
  To cope with cosine similarity, we normalize data to unit vectors and run
  \kde for Euclidean distance. 
  \item \revise{\textbf{\qra}~\cite{DBLP:conf/icdm/Anagnostopoulos15a} and 
  \textbf{\qrb}~\cite{DBLP:journals/tkdd/Anagnostopoulos17} are two query-driven 
  quantized regression models. We use the linear model in \cite{DBLP:conf/icdm/Anagnostopoulos15a} 
  for \qra.}
  %due to the transferability between Euclidean and cosine distance.
  \item \textbf{\lightgbm}~\cite{DBLP:conf/icbsp/WangZZ17} is based on gradient 
  boosting decision trees (CARTs). %and it supports monotonic regression. Here to fair comparison, 
  Each rule in a CART is in the form of $x_i < a$ ($x_i$ is the $i$-th dimension 
  of $\xx$) or $t < b$.
  %We compare with the standard one (\lightgbm) and the one with monotonicity 
  %enforced (\lightgbmm). 
  %xgb~\cite{chen2016xgboost} has the same mechanism with \lightgbm and was shown 
  %to deliver similar accuracy to \lightgbm but in slower speed~\cite{DBLP:conf/sigmod/WangXQ0SWO20}, 
  %so we do not compare it. 
  \item Deep regression models: %We relax the consistency constraint and 
  %We consider the following deep learning models: 
  \textbf{\dnn}, a vanilla feed-forward network; 
  %with four hidden layers of sizes 512, 512, 256, and 256. %
  \textbf{\moe}~\cite{shazeer2017outrageously}, a mixture of expert model with
  sparse activation; 
  %we used 30 expert models, each is an FFN with three hidden
  %layers of sizes 512, 512, and 512. We use top-3 experts for the prediction. %
  \textbf{\rmi}~\cite{kraska2018case}, a hierarchical mixture of expert model; and 
%   We use three levels, and each level has 1, 4, and 8 models. Each model is a
%   feed-forward neural network (FFN) with four hidden layers with sizes 512, 512,
%   256, and 256.
  \textbf{\cardnet}~\cite{DBLP:conf/sigmod/WangXQ0SWO20}, a regression model based 
  on incremental prediction (we enable the accelerated estimation~\cite{DBLP:conf/sigmod/WangXQ0SWO20}). 
  \item Lattice regression models: We adopt \textbf{\dln}~\cite{you2017deep} in this category. 
%   with the architecture of six
%   layers: calibrators, linear embedding, calibrators, ensemble of lattices,
%   calibrators, and linear embedding. 
  \item Clenshaw-Curtis quadrature model: We adopt \textbf{\umnn}~\cite{unconstraintmono}. %in this category. 
  %adopt a feed-forward network with four hidden layers of sizes 512, 512, 256 and 256 to implement the derivative 
  %$\frac{\partial f(\xx, t, \mathcal{D})}{\partial t}$. $f(\xx, t, \mathcal{D})$ is computed by Clenshaw-Curtis Quadrature with learnt derivatives.
  \item Our model is dubbed \textbf{\selnetrp}~\footnote{The source code is available at \cite{source-code}.}. 
  The default setting of $L$ (number of control points) is 50 and $K$ (partition size) 
  is 3. The predefined threshold $\delta_U$ for incremental learning is 20.
%   We use an FFN with three hidden layers to
%   estimate $\Btau$, and an FFN with four hidden layers to estimate $\pp$. The
%   sizes of the first hidden layers of both FFNs are 512, and the sizes of all
%   other hidden layers are 256. The number of control parameters $L$ is $50$, and
%   the random partition number is 3. %
  %%!! mention the impact of #partitions? 
  We also evaluate two \emph{ablated} models: 
  \begin{inparaenum}[(1)]
  \item \textbf{\selnet} is \selnetrp{} without the cover tree partitioning, and 
  \item \textbf{\selnetnotau} is \selnet without the query-dependent feature for 
  control points (disabled by feeding a constant vector into the FFN that 
  generates the $\Btau$ vector). 
  \end{inparaenum}
\end{itemize}
%
% For \is and \kde, we use 2,000 samples to keep the estimation cost
% reasonable. %
% %\kde only supports metric distance functions. 
% %but Cosine distance violates the property of
% %triangle inequality.
% %
% For all the other models, we train them with the same Huber loss on the
% logarithm of the ground truth and prediction; all hyper-parameters are finetuned
% according to the validation set. %
% %
% \dnn{}, \moe{} and \rmi{} cannot directly handle the threshold $t$. We learn a
% non-linear transformation of $t$ into an $m$ dimensional embedding vector, i.e.,
% $\tt = \relu(\ww t)$. %
% Then we concatenate it with $\xx$ as the input to these models. 

% Given a
% training data $(\mathbf{x}, \tau)$, we first adopt a linear embedding
% layer~\cite{you2017deep} to transfer $\tau$ to an embedding $\mathbf{t}$, and
% then concatenate it with $\mathbf{x}$, such that
% $\mathbf{f} = [\mathbf{x}; \mathbf{t}]$, and finally, feed $\mathbf{f}$ into
% models.

%transfer 
%thresholds of Cosine distance to Euclidean distance, and
%adopt \kde to make estimation.

\paragraph{Error Metrics}%
We evaluate Mean Squared Error (\mse), Mean Absolute Percentage Error (\mre), 
and Mean Absolute Error (\mae). 
%on both \valid and \test. 
% They are defined as:
% \begin{align*}
%   & \mathsf{MSE} = \frac{1}{m}\sum_{i=1}^m (\est{y}_i - y_i)^2, & 
%   & \mathsf{MAE} = \frac{1}{m}\sum_{i=1}^m \abs{\est{y}_i - y_i}, \\
%   %& \mathsf{MAPE}  = \frac{1}{m}\sum_{i=1}^m \abs{\frac{\est{y}_i - y_i}{y_i + \varepsilon}}
%   & \mathsf{MAPE}  = \frac{1}{m}\sum_{i=1}^m \abs{\frac{\est{y}_i - y_i}{y_i}}.
% \end{align*}
% where $y_i$ is the ground truth value, $\est{y}_i$ is the estimated value.

%%!! space, and also use y = y+1 smoothing, better than this varepsilon
%%smoothing
%%
% , and
% $\varepsilon = 1 \times 10^{-5}$ is added to \mre{} to avoid division by 0
% error. 

% so that a
% complete comparison among state-of-the-art methods can be
% fully discussed. 

\paragraph{Environment}
Experiments were run on a server with an Intel Xeon E5-2640 @2.40GHz 
CPU and 256GB RAM, running Ubuntu 16.04.4 LTS. 
%Non-deep models were implemented in C++. Deep models were implemented in Python and Tensorflow. 
Models were implemented in Python and Tensorflow.

\newcommand{\mr}[1]{\multirow{2}{*}{#1}}
\newcommand{\mc}[1]{\multicolumn{2}{c}{#1}}
\newcommand{\mcB}[1]{\multicolumn{2}{c|}{#1}}
\newcommand{\mrr}[1]{\multirow{3}{*}{#1}}

\begin{table*}[t]
  \small
  \confversion{\caption{Accuracy (\mse and \mae measured in $10^5$ and $10^2$, respectively).}}
  \fullversion{\caption{Accuracy (\mse and \mae measured in $10^5$ and $10^2$, respectively).}}
  \label{exp:tab:accuracy}  
  \centering
  \resizebox{\textwidth}{!}{  
  \begin{tabular}{l||r|r|r|r|r|r|r|r|r|r|r|r|r|r|r|r|r|r}
    \hline
    \mr{Model} & \multicolumn{3}{c|}{\fasttext} & \multicolumn{3}{c|}{\glove} & \multicolumn{3}{c|}{\face} & \multicolumn{3}{c|}{\youtube} & \multicolumn{3}{c|}{\deep} & \multicolumn{3}{c}{\sift} \\ \cline{2-19} 
            & \mse & \mae & \mape & \mse & \mae & \mape & \mse & \mae & \mape & \mse & \mae & \mape & \mse & \mae & \mape & \mse & \mae & \mape \\ \hline\hline
    \rs*        &  22.38  & 7.45  & 1.40 & 34.85 & 8.71 & 1.09 & 28.64 & 8.09 & 1.30 & 2.95 & 1.89 & 0.88 & 84732.32 & 725.21 & 1.26 & 30317437.60 & 9496.55 & 0.98 \\ %\hline
    \is*        & -      & -     & -    & -     & -    & -    & 104.58 & 14.25 & 1.25 & 2.85 & 1.83 & 0.76 & 50242.10 & 314.12 & 0.97 & 32049511.88 & 10612.42 & 0.92 \\ %\hline
    \kde*       & 21.46  & 6.57  & 1.28 & 37.52 & 8.93 & 0.87 & 36.43 & 8.42 & 1.02 & 2.93 & 1.90 & 0.70 & 39497.24 & 298.10 & 0.85 & 27651109.31 & 9581.89 & 0.91 \\ %\hline
    \qra        &   41.79    &  8.95 & 1.02 & 50.13 & 9.21 & 0.99 &  74.35 &  11.60 & 1.06    & 3.42     &     2.01 &  0.71 & 37054.11 & 292.16 & 0.78 & 28077423.18 & 9953.34 & 0.94 \\ %\hline
    \qrb        &   34.35     & 8.03 & 0.97 & 42.47 & 9.01 & 0.86 & 42.94  & 9.05  & 0.89  &   2.74   & 1.85 & 0.55 & 31485.85 & 273.22 & 0.67 & 22048511.27 & 8542.73 & 0.71 \\ %\hline
    \lightgbm   & 98.77  & 9.56  & 1.04 & 72.11 & 10.87 & 0.89 & 101.29 & 9.51  & 0.45 & 4.01 & 2.00 & 0.52 & 44036.45 & 301.88 & 0.85 & 23849121.36 & 9005.17 & 0.75 \\ %\hline
    %\lightgbmm* & 101.22 & 9.84  & 1.20 & 76.31 & 11.10 & 0.92 & 114.62 & 10.46 & 0.49 & 5.43 & 2.19 & 0.65 & 48110.28 & 310.41 & 0.89 & 29840392.08 & 9244.23 & 0.77 \\ %\hline
    \dnn        & 63.54  & 11.25 & 1.33 & 52.31 & 9.39 & 0.91 & 110.77 & 17.14 & 0.89 & 2.78 & 1.77 & 0.51 & 20454.11 & 192.13 & 0.69 & 24465910.26 & 7144.68 & 0.55 \\ %\hline  
    \moe        & 45.90 & 8.50  & 0.91 & 30.14 & 7.05 & 0.91 & 21.25  & 4.32  & 0.30 & 1.58 & 1.59 & 0.53 & 18068.93 & 170.51 & 0.65 & 14750194.30 & 6327.47 & 0.40 \\ %\hline
    \rmi        & 26.16  & 6.10  & 0.87 & 29.32 & 6.89 & 0.74 & 22.16  & 6.07  & 0.35 & 1.77 & 1.62 & 0.55 & 9498.21 & 116.54 & 0.67 & 8906108.00 & 4650.29  & 0.42 \\ %\hline
    \cardnet*   & 25.67  & 6.16  & 0.90 & 27.05 &  6.19 & 0.78 & 13.67  & 4.08  & 0.27 & 1.41 & 1.44 & 0.48 & 9230.48 & 117.37 & 0.67 & 7248693.54 & 4851.43 & 0.40 \\ %\hline
    \dln*       & 77.50  & 11.56 & 1.53 & 52.26 & 10.27 & 0.89 & 82.35 & 11.85 & 0.97 & 2.94 & 1.92 & 0.69 & 58291.42 & 353.08 & 0.94 & 23059384.16 & 8058.49 & 0.51 \\ %\hline
    \umnn*      & 33.26  & 7.20  & 0.92 & 33.50 & 7.98 & 0.86 & 16.75  & 4.70  & 0.36 & 2.06 & 1.69 & 0.49 & 10603.68 & 131.04 & 0.73 & 10201332.32 & 5443.31 & 0.43 \\ \hline
    \selnetrp*  & \textbf{7.87}  & \textbf{3.56} & \textbf{0.76} & \textbf{9.17} & \textbf{3.83} & \textbf{0.68} & \textbf{4.96} & \textbf{2.43} & \textbf{0.23} & \textbf{0.72} & \textbf{1.13} & \textbf{0.36} & \textbf{2243.42} & \textbf{51.92} & \textbf{0.51} & \textbf{1464247.70} & \textbf{1406.63} & \textbf{0.23} \\
    \hline
  \end{tabular}
  }
\end{table*}

\subsection{Accuracy}
We report accuracies in Table~\ref{exp:tab:accuracy}, where monotonic models 
are marked with *, and best values are marked in boldface. 
%\begin{itemize}
%  \item 
Our model, \selnetrp{}, consistently outperforms 
existing models. It achieves substantial error reduction against the best of 
state-of-the-art methods, in all the three error metrics and all the settings. 
\revise{Compare to the runner-up model on each dataset, the improvement is 
2.0 -- 5.0 times in \mse{}, 1.3 -- 3.3 times in \mae{}, and 1.2 -- 1.7 times 
in \mape{}, and is more significant on larger datasets.}

We examine each category of models. 
\revise{We start with the sampling-based methods. \kde works better than \rs{} and \is{} in most 
settings. In fact, \kde's performance even outperforms some deep learning regression 
based methods in a few cases (e.g., \mse on \fasttext). 
% One possible reason
% is that it focuses on the metric space and escapes the curse of dimensionality.
Among non-deep learning models, these is no best model across all the datasets, though 
\qrb prevails on more datasets than others.} 
%\lightgbmm is worse than \lightgbm. 
%This indicates that the monotonic constraint, albeit better interpretability, 
%decreases the performance of \lightgbm. 
Among the deep learning models other than ours, \cardnet is generally the best 
thanks to its incremental prediction for each threshold interval. 
%\moe is not good at large selectivities (\mse on \fasttext). 
The performance of \dln is mediocre. The main reason is analyzed in Section~\ref{sec:compare-lattice-regression}. 
% its inherent limitation to model complex regression problems. 
% It considers a narrow class of multi-dimensional monotonic functions, while in 
% our problem the distance function is a one-dimensional (i.e., $t$) monotonic function.
The accuracy of \umnn, which uses the same integral points for different queries, 
though better than \dln, still trails behind ours by a large margin. 

\fullversion{
\subsection{Consistency Test}

% mono test
%\vspace{-2mm}
\begin{table}[t]
  \small
  \caption{Empirical monotonicity (\%) on \face.}
  \label{exp:tab:mono}  
  \centering
  \begin{tabular}{c c c c c}
    \rs*      & \is*    & \kde*  & \qra       & \qrb  \\
    100       & 100     & 100    &   85.39  & 84.86       \\ \hline
    \lightgbm & \dnn    & \moe   & \rmi       &       \\
    86.34     & 78.22   & 94.82  & 90.48      &       \\ \hline
    \cardnet* & \dln*   & \umnn* & \selnetrp* &       \\
    100       & 100     & 100    & 100        &
  \end{tabular}
\end{table}
%\vspace{-2mm}

We compute the empirical monotonicity measure~\cite{daniels2010monotone} 
and show the results in Table~\ref{exp:tab:mono}. The measure is the 
percentage of estimated pairs that violate the monotonicity, averaged 
over 200 queries. For each query, we sampled 100 thresholds, which form 
$\binom{100}{2}$ pairs. A low score indicates more inconsistent estimates. 
As expected, models without consistency guarantee cannot produce 100\% 
monotonicity. 
%indeed do not process this property.

% $\mathsf{mono}(\mathcal{T}) = \frac{\text{\#Monotone
%     pairs}(\mathcal{T})}{\text{\#Comparable
%     pairs}(\mathcal{T})}$

% We verify the monotonicity in 

% We generate $\mathcal{T}$ by randomly sampling 200
% queries from the testing data of \facecos and constructing an
% arithmetic sequence
% of 101 thresholds in the range $[0, 1]$ for each query. One pair is
% defined as $(\xx, t, t')$, where $\xx$ is a query,
% and $t$, $t'$ ($t \leq t'$) are adjacent thresholds
% in the sequence, e.g., 0.01 and 0.02. The experimental results show
% non-monotonic models (i.e., \dnn, \rmi, \moe) do not achieve very good
% consistency, especially \dnn. Our models have the consistency and also
% achieve better performance.
}

\subsection{Ablation Study}\label{sec:exp-abalation}

\begin{table}[t]
  \small
  \caption{Ablation study.}
  \label{exp:tab:ablation}  
  \centering
  \begin{tabular}{l||r|r|r|r}
    \hline
    Dataset         & Model        & \mse ($\times 10^5$) & \mae ($\times 10^2$) & \mre \\ \hline\hline
    \mrr{\fasttext} & \selnetrp    & \textbf{7.87} & \textbf{3.56} & \textbf{0.76} \\%\hline
                    & \selnet      & 12.63         & 4.37          & 0.81          \\ %\hline
                    & \selnetnotau & 39.59         & 8.72          & 2.90          \\ \hline
    \mrr{\glove}    & \selnetrp    & \textbf{9.17} & \textbf{3.83} & \textbf{0.68} \\ %\hline
                    & \selnet      & 22.43         & 5.82          & 0.70          \\ %\hline
                    & \selnetnotau & 32.59         & 6.92          & 0.90          \\ \hline 
    \mrr{\face}     & \selnetrp    & \textbf{4.96} & \textbf{2.43} & \textbf{0.23} \\
                    & \selnet      & 5.31          & 2.92          & 0.24          \\
                    & \selnetnotau & 16.02         & 4.65          & 0.37          \\ \hline
    \mrr{\youtube}  & \selnetrp    & \textbf{0.72} & \textbf{1.13} & \textbf{0.36} \\ %\hline
                    & \selnet      & 0.90          & 1.20          & 0.39          \\ %\hline
                    & \selnetnotau & 1.65         & 1.59          & 0.53          \\ \hline
    \mrr{\deep}     & \selnetrp    & \textbf{2243.42} & \textbf{51.92} & \textbf{0.51} \\ %\hline
                    & \selnet      & 5861.43       & 72.18         & 0.58          \\ %\hline
                    & \selnetnotau & 9012.57       & 101.42        & 0.71          \\ \hline
    \mrr{\sift}     & \selnetrp    & \textbf{1464247.70} & \textbf{1406.63} & \textbf{0.23} \\ %\hline
                    & \selnet      & 4958113.22    & 2911.86       & 0.27          \\ %\hline
                    & \selnetnotau & 6904808.25    & 3855.53       & 0.54          \\ \hline
  \end{tabular}
\end{table}

%\paragraph{\selnet v.s. \selnetrp}
%
% The main difference between the two models is whether cover tree partition is
% used. 
Table~\ref{exp:tab:ablation} shows that the partitioning (\selnetrp v.s. \selnet) 
improves \mse, \mae, and \mape by up to 3.4, 2.1, and 1.2 times, respectively, and 
the effect is more remarkable on large datasets. This is because each model deals 
with a subset of the dataset for better fit and the ground truth label values for 
each model are reduced, which makes it easier to fit our piecewise linear function 
with the same number of control points, as the value function is less steep. 
Using query-dependent control points (\selnet v.s. \selnetnotau) also has a 
significant impact on accuracy across all the settings and all the error metrics. 
%When query-dependent control points are employed to fit the selectivity curve, 
The improvements in \mse, \mae, and \mape are up to 3.1, 2.0, and 3.6 times, 
respectively. %

\subsection{Estimation Time}

% In the selectivity estimation problem,  our models (and other machine learning models) are trained offline or in the preprocessing step. We argue that this is a reasonable setting because many DB techniques require a preprocessing step to assist the online processing, such as constructing index structures. Thus, in the query processing step, our model just makes inference for each query, and does not need to learn parameters. 
Table~\ref{tab:time} reports the estimation times of the competitors. 
We also report the time of running a state-of-the-art selection 
algorithm (\covertree~\cite{fastercovertree}) to obtain the exact 
selectivity. 
\revise{All the models except \is are at least one order of magnitude 
faster than \covertree, and the gaps increase to three orders of 
magnitude on \deep and \sift. 
Our model is on a par with other deep learning models (except \dnn) 
and faster than sampling and quantized regression methods.} 
% Compared with DB approaches, i.e., \is and \kde, our models have 
% faster prediction time, e.g., at least 2.8 times faster than the 
% best of DB approaches. 
% Although \dnn is very fast due to its simple model structure, 
% its accuracy is much worse than ours, as we have witnessed in 
% Table~\ref{exp:tab:accuracy}. 
% Among ML models, \dnn is the fastest due to
% its simple model structure. Our models become the runner-up, and 
% their speeds are faster than \rmi, \moe \dln and \umnn. 

\begin{table}[t]
  \small
  \caption{Average estimation time (milliseconds).}
  \label{tab:time}  
  \centering
  \resizebox{\linewidth}{!}{
  \begin{tabular}{l||r|r|r|r|r|r}
    \hline
    Model         & \fasttext & \glove & \face & \youtube & \deep & \sift \\ \hline \hline
    \covertree    & 8.14      & 8.85   & 9.65  & 6.11     & 214   & 395   \\
    %\faiss        &           &        &       &          &       &       \\
    \rs*          & 0.46      & 0.51   & 0.49  & 0.52     & 2.54  & 4.72  \\
    \is*          & -         & -      & 1.08  & 2.35     & 4.97  & 6.81  \\
    \kde*         & 0.79      & 0.68   & 0.59  & 0.94     & 1.48  & 2.05  \\
    \qra          & 0.86      & 0.98   & 0.97  & 1.03     & 2.21  & 2.82  \\
    \qrb          & 0.79      & 0.99   &  0.95 & 1.10     & 2.32  & 2.91  \\
    \lightgbm     & 0.28      & 0.30   & 0.18  & 0.52     & 0.26  & 0.26  \\
    %\lightgbmm*   & 0.28      & 0.31   & 0.19  & 0.50     & 0.25  & 0.26  \\
    \dnn          & \textbf{0.07} & \textbf{0.10} & \textbf{0.03} & \textbf{0.16} & \textbf{0.11} & \textbf{0.10} \\
    \moe          & 0.36      & 0.33   & 0.27  & 0.49     & 0.29  & 0.33  \\
    \rmi          & 0.34      & 0.38   & 0.25  & 0.47     & 0.27  & 0.30  \\
    \cardnet*     & 0.19      & 0.26   & 0.14  & 0.31     & 0.22  & 0.28  \\ 
    \dln*         & 0.83      & 0.69   & 0.65  & 1.22     & 0.64  & 0.80  \\
    \umnn*        & 0.39      & 0.32   & 0.24  & 0.52     & 0.26  & 0.32  \\ \hline
    \selnetrp*    & 0.35      & 0.31   & 0.24  & 0.51     & 0.29  & 0.36  \\
    %\selnet*      & 0.17  & \textbf{0.16}    & \textbf{0.10}  & \textbf{0.28} \\
    %\selnetnotau* & \textbf{0.16}      & 0.18 & 0.11 & 0.29         \\ %\hline
    \hline
  \end{tabular}
  }
\end{table}
%\vspace{-4mm}

\begin{table} [t]
  \small%\scriptsize
  \caption{Training time (hours).}
  \label{tab:traintime}
  \centering
  \resizebox{\linewidth}{!}{
  \begin{tabular}{l||r|r|r|r|r|r}
    \hline
    Model         & \fasttext & \glove & \face & \youtube & \deep & \sift \\ \hline \hline
    \kde*         & \textbf{1.1} & \textbf{1.5} & \textbf{0.7} & \textbf{0.8} & 2.5 & 3.6 \\
    \qra          & 1.8       & 2.1    & 1.5   & 1.2      & 3.9   & 4.6   \\
    \qrb          & 1.5       & 2.0    & 1.6   & 1.2      & 3.6   & 4.7   \\
    \lightgbm     & 2.1       & 1.9    & 2.2   & 2.1      & \textbf{1.9} & \textbf{2.1} \\
    %\lightgbmm*   & 1.4       & 1.8    & 1.2   & 1.1      & 1.4   & 1.4   \\
    \dnn          & 2.9       & 2.1    & 2.8   & 2.9      & 2.5   & 2.9   \\
    \moe          & 4.9       & 5.4    & 4.9   & 4.7      & 4.3   & 4.4   \\
    \rmi          & 5.4       & 5.6    & 4.8   & 5.3      & 4.6   & 4.8   \\
    \cardnet*     & 3.8       & 3.9    & 3.3   & 3.2      & 3.5   & 3.8   \\    
    \dln*         & 6.9       & 7.1    & 6.0   & 6.5      & 6.3   & 6.4   \\
    \umnn*        & 5.5       & 5.7    & 4.9   & 5.2      & 5.4   & 4.6   \\ \hline
    \selnetrp*    & 6.0       & 5.5    & 5.2   & 5.6      & 5.2   & 5.0   \\
    %\selnet*      & 3.6  & 3.5 & 3.1  & 3.0              \\
    %\selnetnotau* & 3.6  & 3.7 & 3.0 & 2.9        \\ %\hline
    \hline
  \end{tabular}
  }
\end{table}

\fullversion{
\begin{figure} [t]
  \centering
  \subfigure[\mse, \fasttext]{
    \includegraphics[width=0.46\linewidth]{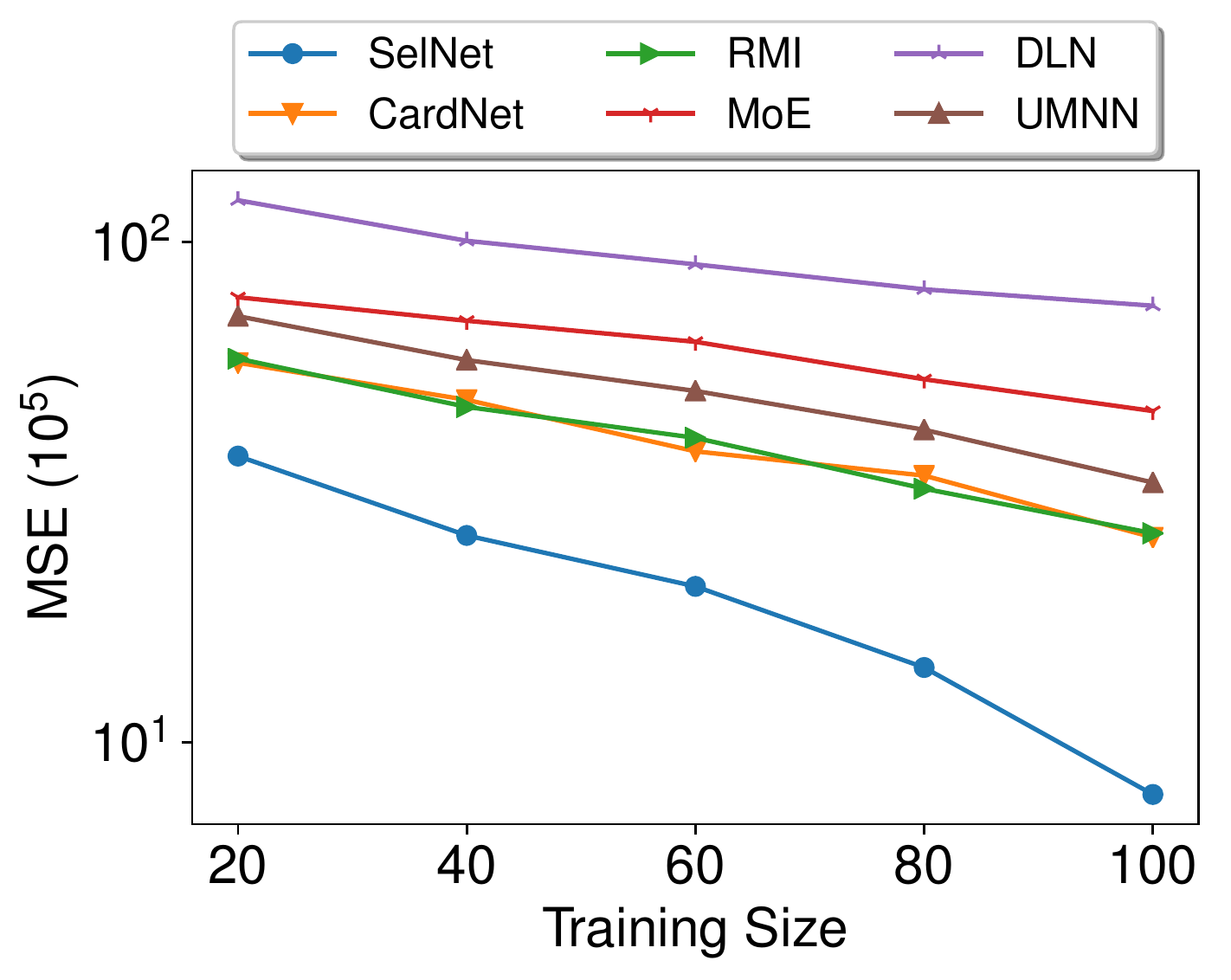}
    \label{fig:exp-trainsize-fasttext}
  }
  \subfigure[\mse, \glove]{
    \includegraphics[width=0.46\linewidth]{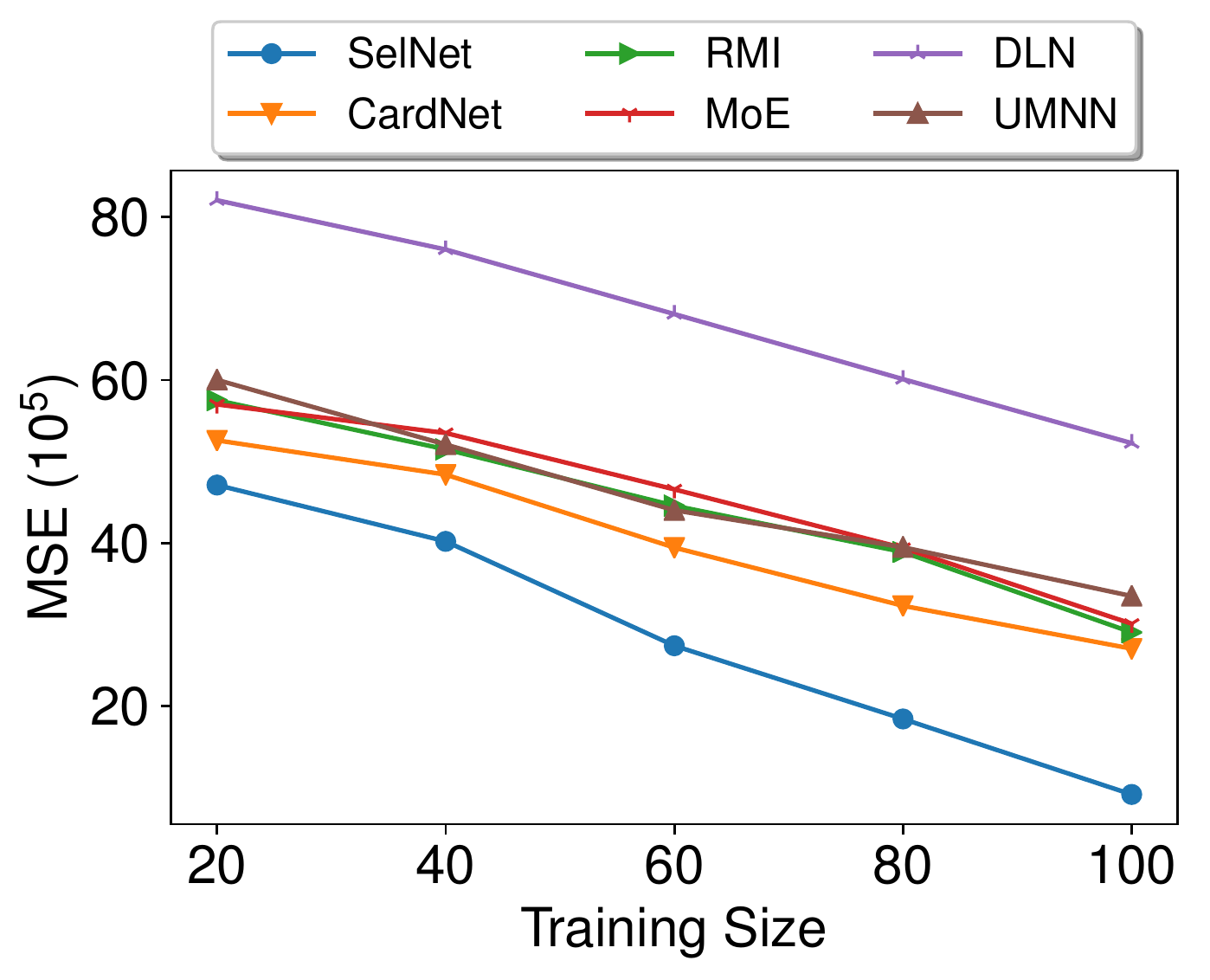}
    \label{fig:exp-trainsize-glove}
  }
  \subfigure[\mse, \face]{
    \includegraphics[width=0.46\linewidth]{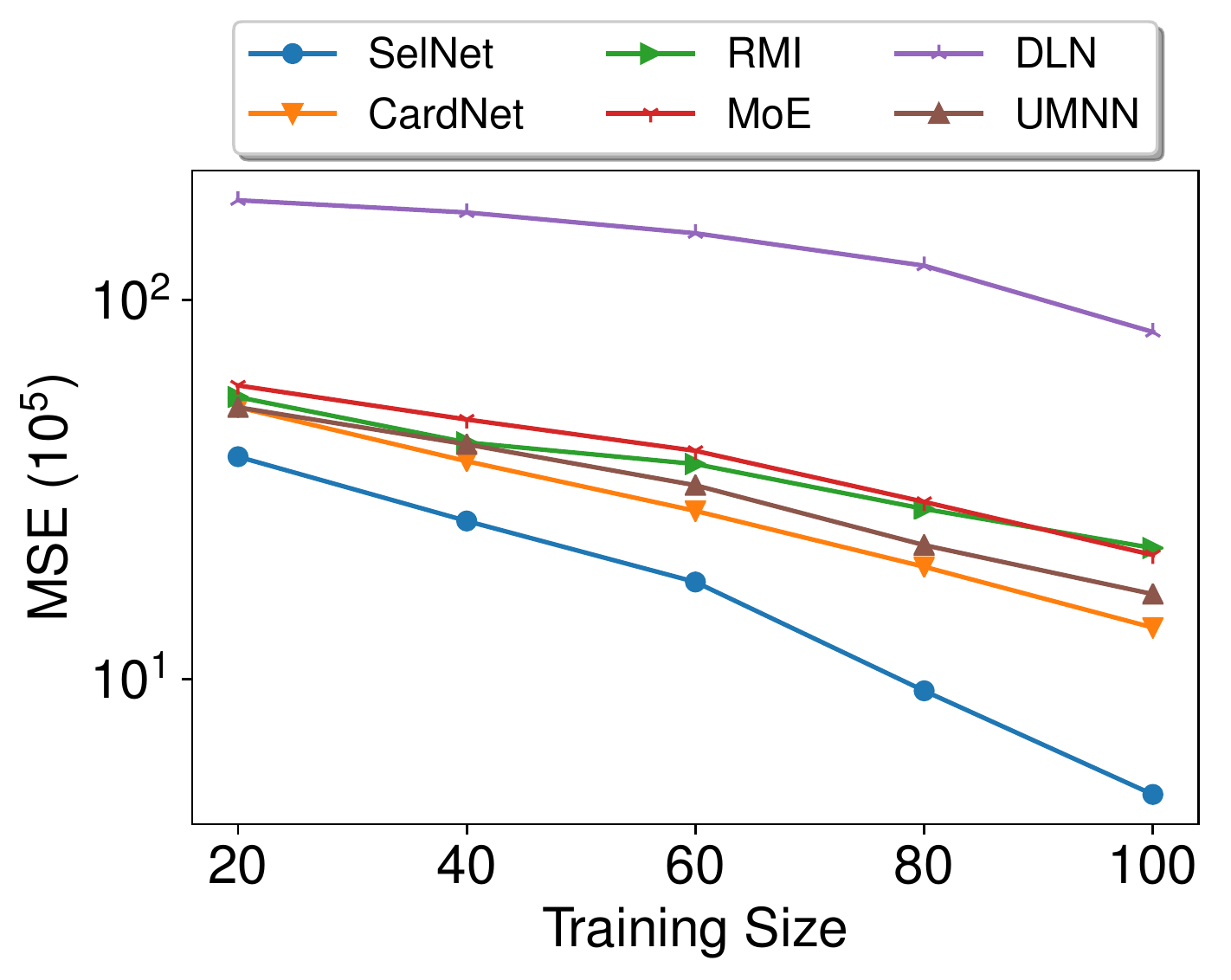}
    \label{fig:exp-trainsize-face}
  }
  \subfigure[\mse, \youtube]{
    \includegraphics[width=0.46\linewidth]{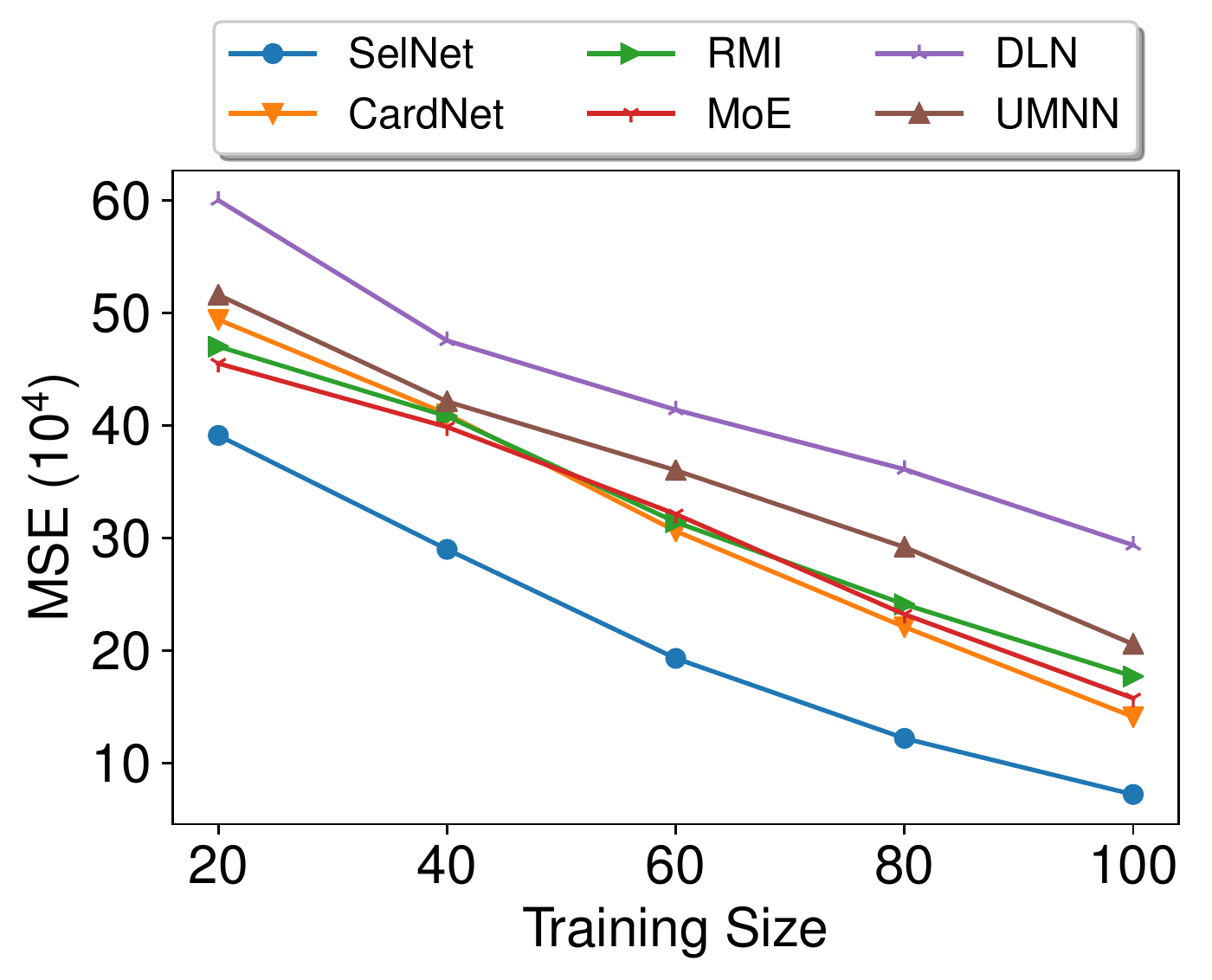}
    \label{fig:exp-trainsize-youtube}
  }
  \subfigure[\mse, \deep]{
    \includegraphics[width=0.46\linewidth]{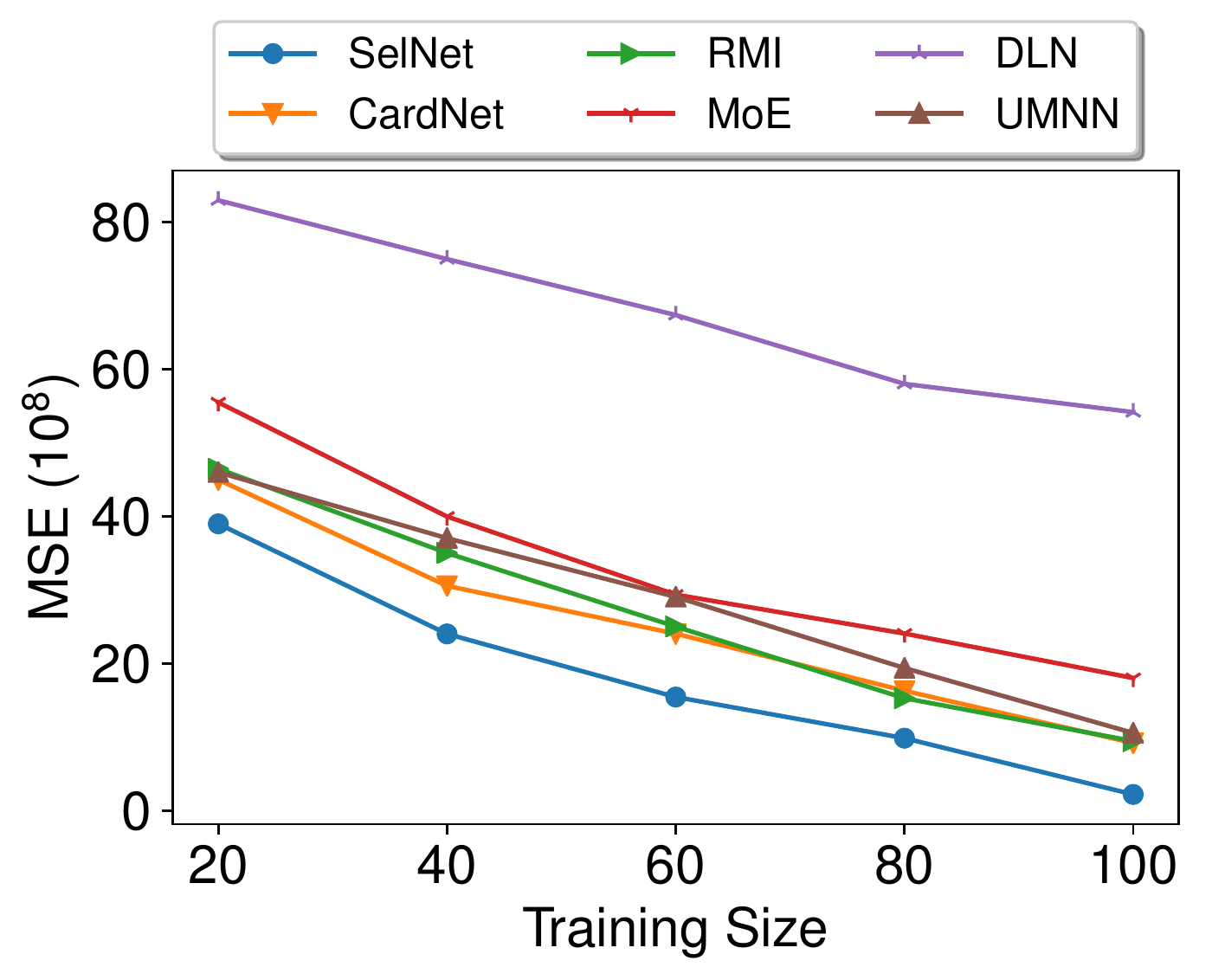}
    \label{fig:exp-trainsize-deep}
  }
  \subfigure[\mse, \sift]{
     \includegraphics[width=0.46\linewidth]{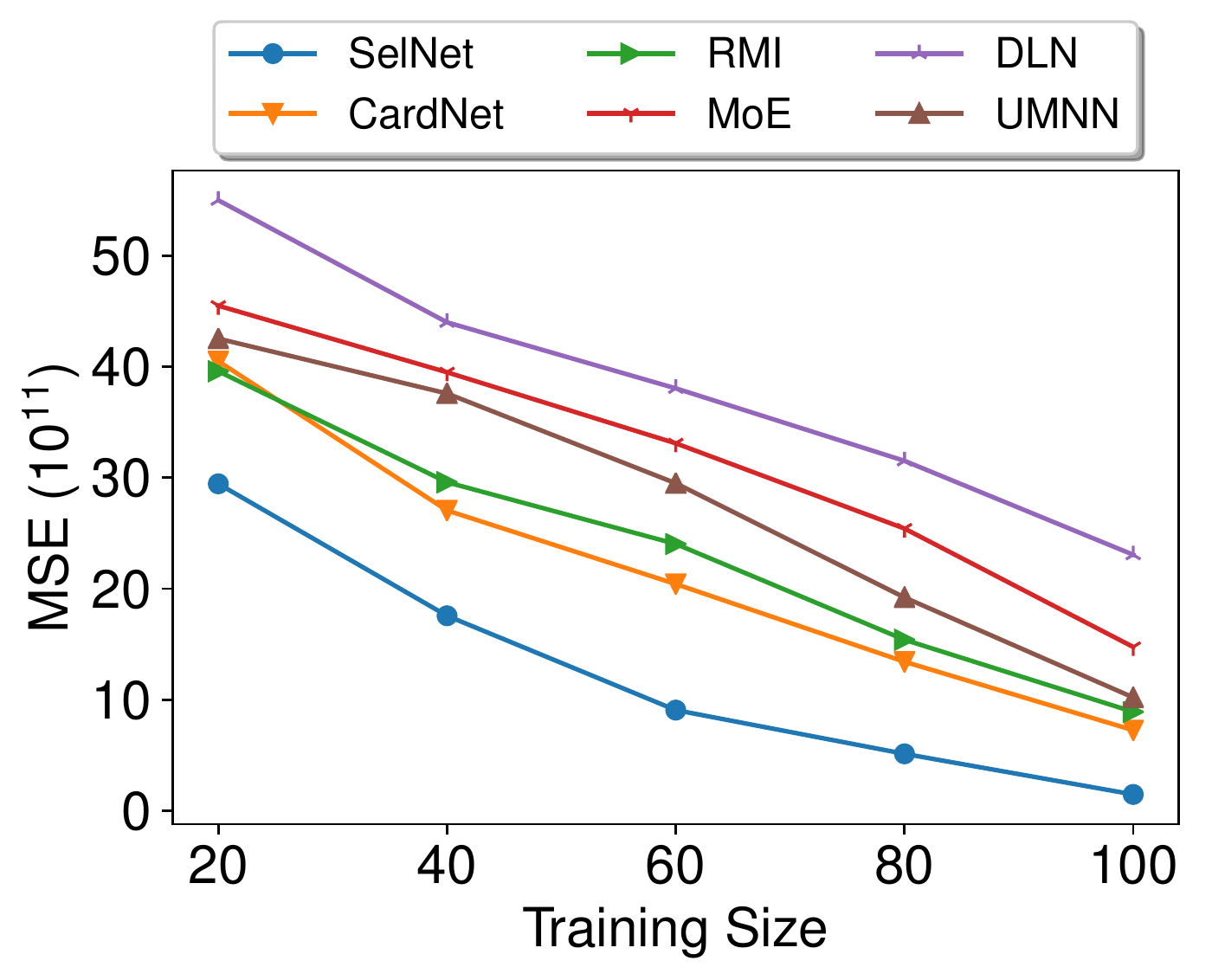}
    \label{fig:exp-trainsize-sift}
  }  
  \caption{Varying training data size.}
  \label{fig:exp-trainsize}
\end{figure}
}

\subsection{Training}
Table~\ref{tab:traintime} shows the training times. Non-deep models are faster to train. 
Our models spend 5 -- 6 hours, similar to other deep models. 
\fullversion{In Figure~\ref{fig:exp-trainsize}, we show the performances, measured by 
\mse, of the deep learning models by varying the scale of training examples from 
20\% to 100\% of the original training data. All the models perform worse with fewer 
training data, but our models are more robust, showing moderate accuracy loss. 
}

\subsection{Data Update}
%\vspace{-1.5mm}
\begin{figure}[t]
  \centering
  \subfigure[\mse]{
    \includegraphics[width=0.46\linewidth]{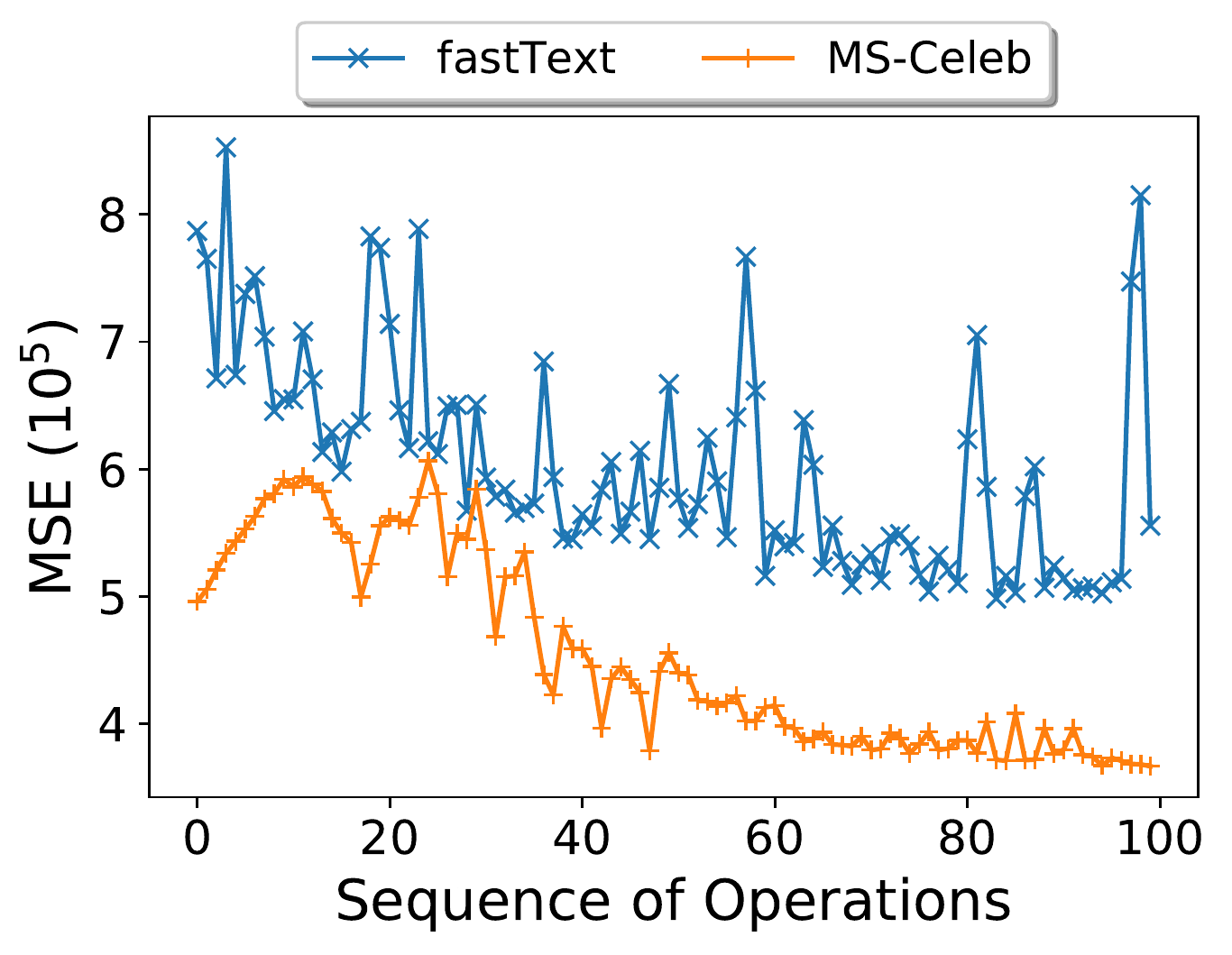}
    \label{fig:exp-update-1}
  }
  \subfigure[\mape]{
    \includegraphics[width=0.46\linewidth]{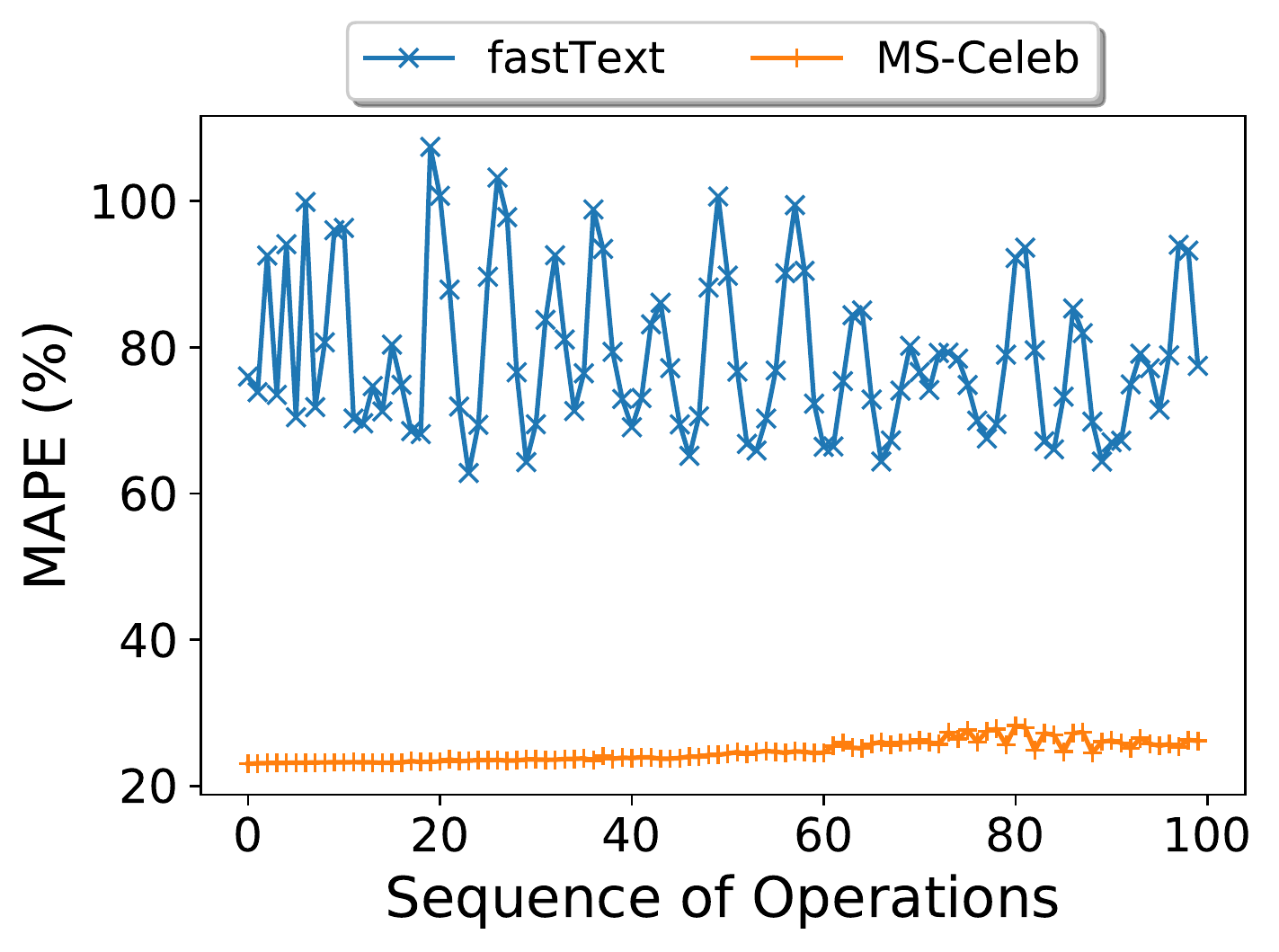}
    \label{fig:exp-update-2}
  }  
  \caption{Data update.}
  \label{fig:exp-update}
\end{figure}
%\vspace{-0.5mm}

We generate a stream of 100 update operations, each with an insertion
or deletion of 5 records on \fasttext and \face, to evaluate 
our incremental learning technique. 
% , and adopt incremental learning in Section~\ref{sec-update} for \selnetrp in \facecos{}
% and \fasttextcos datasets. 
Figure~\ref{fig:exp-update} plots how \mse and \mape change with the 
stream. The general trend is that the \mse is decreasing when there 
are more updates, while \mape fluctuates or keeps almost the same. 
Such difference is caused by the change of labels (i.e., true 
selectivities) in the stream. Nonetheless, the result indicates that 
incremental learning is able to keep up with the updated data. Besides, 
\selnetrp only spends 1.5 -- 2.0 minutes for each incremental learning, 
showcasing its speed to cope with updates.

\subsection{Evaluation of Hyper-Parameters}

%\paragraph{Varying the Number of Partitions}
\label{sec:exp-control-points}

\begin{table}[t]
  \small
  \caption{Varying number of control points on \fasttext.}
  \label{exp:tab:interpolation}  
  \centering
  \begin{tabular}{l || r r r r}
    \mr{Error Metric} & \multicolumn{4}{c}{Number of Control Points}\\\cline{2-5}    
                          & 10 & 50  & 90 & 130  \\ \hline\hline
    \mse ($\times 10^5$)  & 13.06 & 7.87  & 7.93  & 10.47 \\
    \mae ($\times 10^2$)  & 4.85 & 3.56  & 3.56  & 3.92 \\
    \mape                 & 0.87 & 0.76 & 0.76 & 0.79
  \end{tabular}
\end{table}

%The main hyper-parameters in our models are the number of control points $L$ and the 
%partition size $K$. %in the cover tree partitioning. 
%Due to the space limitation, we do not show the effect of other hyper-parameters. 
%because they are less important than $L$
%and $K$.
%We make experiments in the validation data as follows.
%The number of control points is related with the final performance. 

Table~\ref{exp:tab:interpolation} shows the accuracy when we vary the number of control 
points $L$ on \fasttext. 
A small value leads to underfitting towards the curve of thresholds, while a large 
value increases the learning difficulty. $L = 50$ achieves the best performance. 

%\subsection{Evaluation of Partitioning}
%\vspace{-4mm}
\begin{table}[t]
  \small
  \caption{Varying partition size on \fasttext.}
  \label{exp:tab:partition}  
  \centering
  \begin{tabular}{l || r r r r}
    \mr{Error Metric} & \multicolumn{4}{c}{Partition Size}\\\cline{2-5}    
                          & 1 & 3    & 6    & 9   \\ \hline\hline
    \mse ($\times 10^5$) & 12.63 & 7.87  & 6.82  & 6.75 \\
    \mae ($\times 10^2$) & 4.37 & 3.56  & 3.36  & 3.11 \\
    \mape                & 0.81 & 0.76 & 0.77 & 0.74 \\
    Estimation Time (ms)        & 0.16 & 0.35 & 0.79 & 1.24
  \end{tabular}
\end{table}
%\vspace{-4mm}
% \mse ($\times 10^5$)         & 784732  & 678917  & 672865  \\
% \mae          & 355  & 330  & 308  \\
% \mape         & 0.75  & 0.77  &  0.74

\begin{table}[t]
  \small
  \caption{Varying partitioning method on \fasttext.}
  \label{exp:tab:partmethod}  
  \centering
  \begin{tabular}{l || r r r}
    Error Metric & CT (3) & RP (3) & KM (3) \\ \hline\hline
    \mse ($\times 10^5$)  & 7.87 & 8.02 & 9.14 \\
    \mae ($\times 10^2$)  & 3.56 & 3.57  & 3.64   \\
    \mape                 & 0.76 & 0.78 & 0.79
  \end{tabular}
\end{table}

Table~\ref{exp:tab:partition} reports the accuracy when we vary the partition size 
$K$ on \fasttext. There is no partitioning when $K = 1$. We observe that the 
partitioning is useful, but the improvement is small when partition size exceeds 3, 
and estimation time also substantially increases. This means a small partition 
size ($K = 3$) suffices to achieve good performance. 
For partitioning strategy, we compare cover tree partitioning (CT) with random 
partitioning (RP) and $k$-means partitioning (KM) in Table~\ref{exp:tab:partmethod}. 
CT delivers the best performance. KM is the worst because it tends to cause 
imbalance in the partition. 
%Furthermore, when $t$ is small, CT is more robust than RP, because CT is able to filter many invalid regions.

\fullversion{
\subsection{Generalizability}

\revise{To show the generalizability of our model, we evaluate the performance 
on the queries that significantly differ from the records in the training data. 
To prepare such queries, we first perform a $k$-means clustering on $\mathcal{D}$.  
We randomly sample 10,000 query objects from $\mathcal{D}$ (excluding the 
queries used for training) and add Gaussian noise~\cite{zhang2018word}. Then we 
pick the top-2,000 ones having the largest sum of squared distance to the $k$ 
centroids. Figure~\ref{fig:exp-unseen} show the performances of \kde, \rmi, 
\cardnet, and \selnetrp on \deep and \sift, measured by \mse. The queries are 
grouped by selectivity range. In each selectivity group, \selnetrp consistently 
outperforms the other models, and the advantage is around one order of magnitude. 
This result demonstrates that our model generalizes well for out-of-dataset 
queries.}

\begin{figure}[t]
  \centering
  \subfigure[\deep]{
    \includegraphics[width=0.46\linewidth]{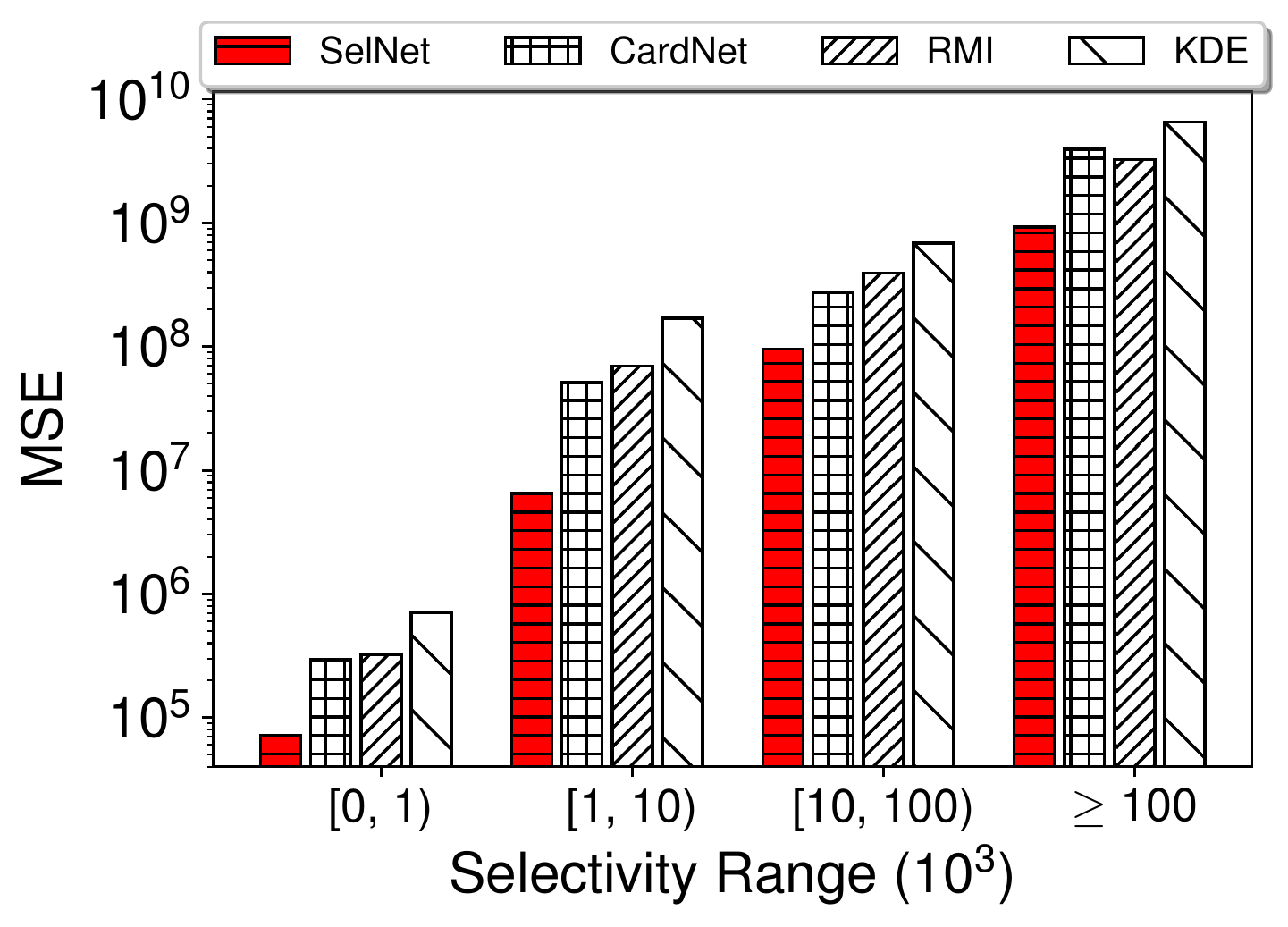}
    \label{fig:exp-unseen-1}
  }
  \subfigure[\sift]{
    \includegraphics[width=0.46\linewidth]{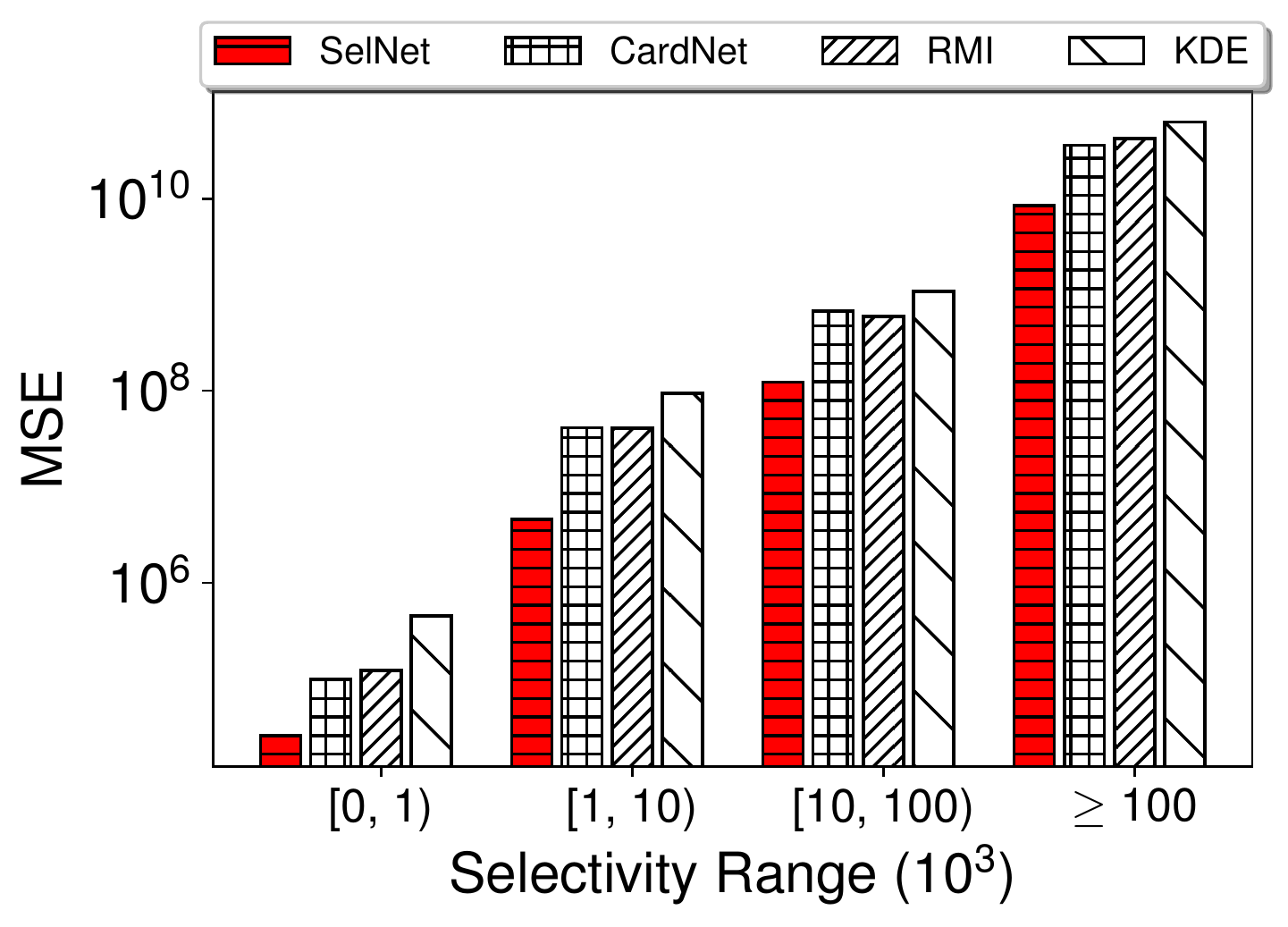}
    \label{fig:exp-unseen-2}
  }  
  \caption{Generalizability.}
  \label{fig:exp-unseen}
\end{figure}
}

\revise{
\subsection{Performance in Semantic Search}

%\vspace{-1.5mm}
\begin{figure}[t]
  \centering
  \subfigure[\aminer]{
    \includegraphics[width=0.46\linewidth]{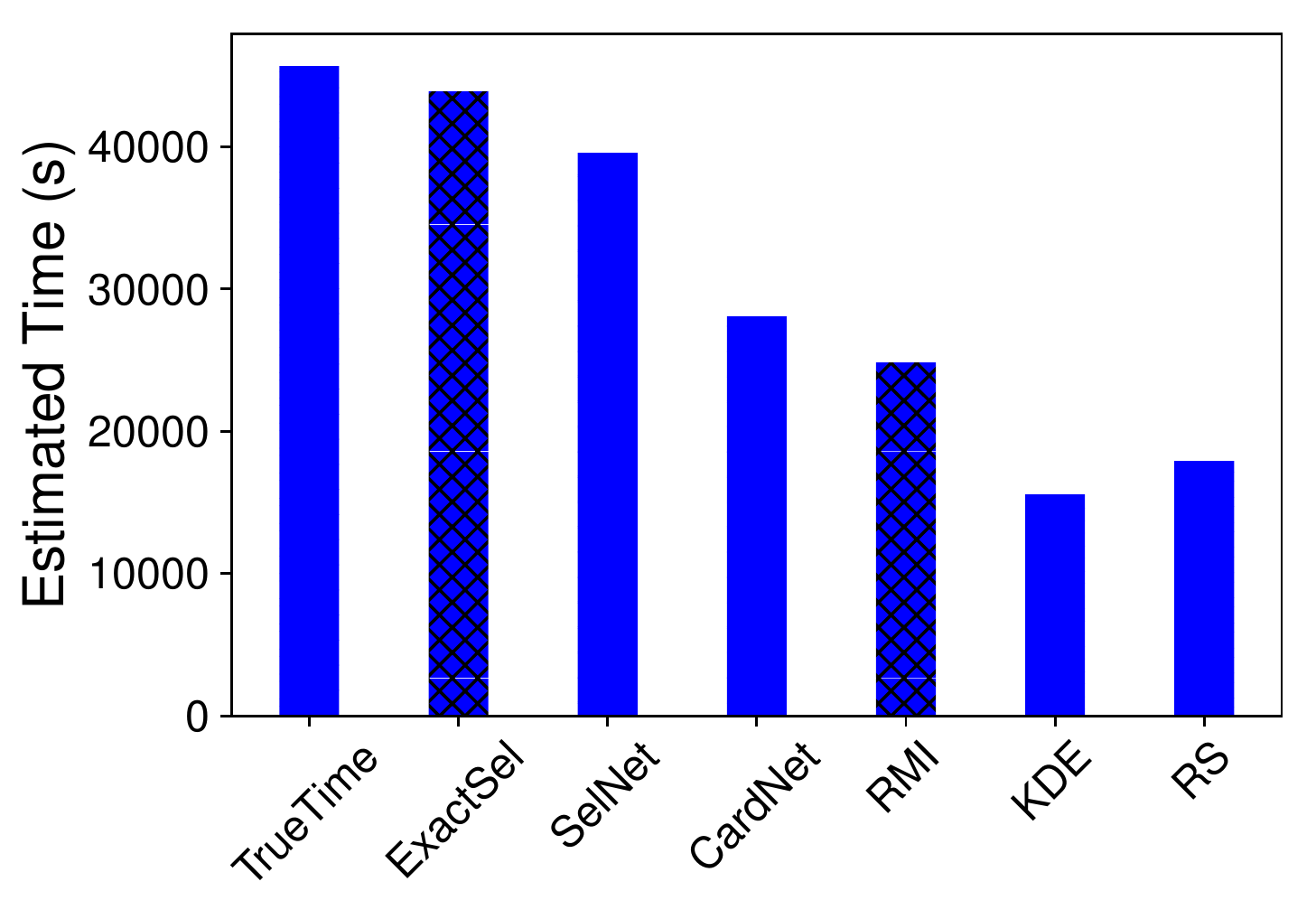}
    \label{fig:exp-application-1}
  }
  \subfigure[\quora]{
    \includegraphics[width=0.46\linewidth]{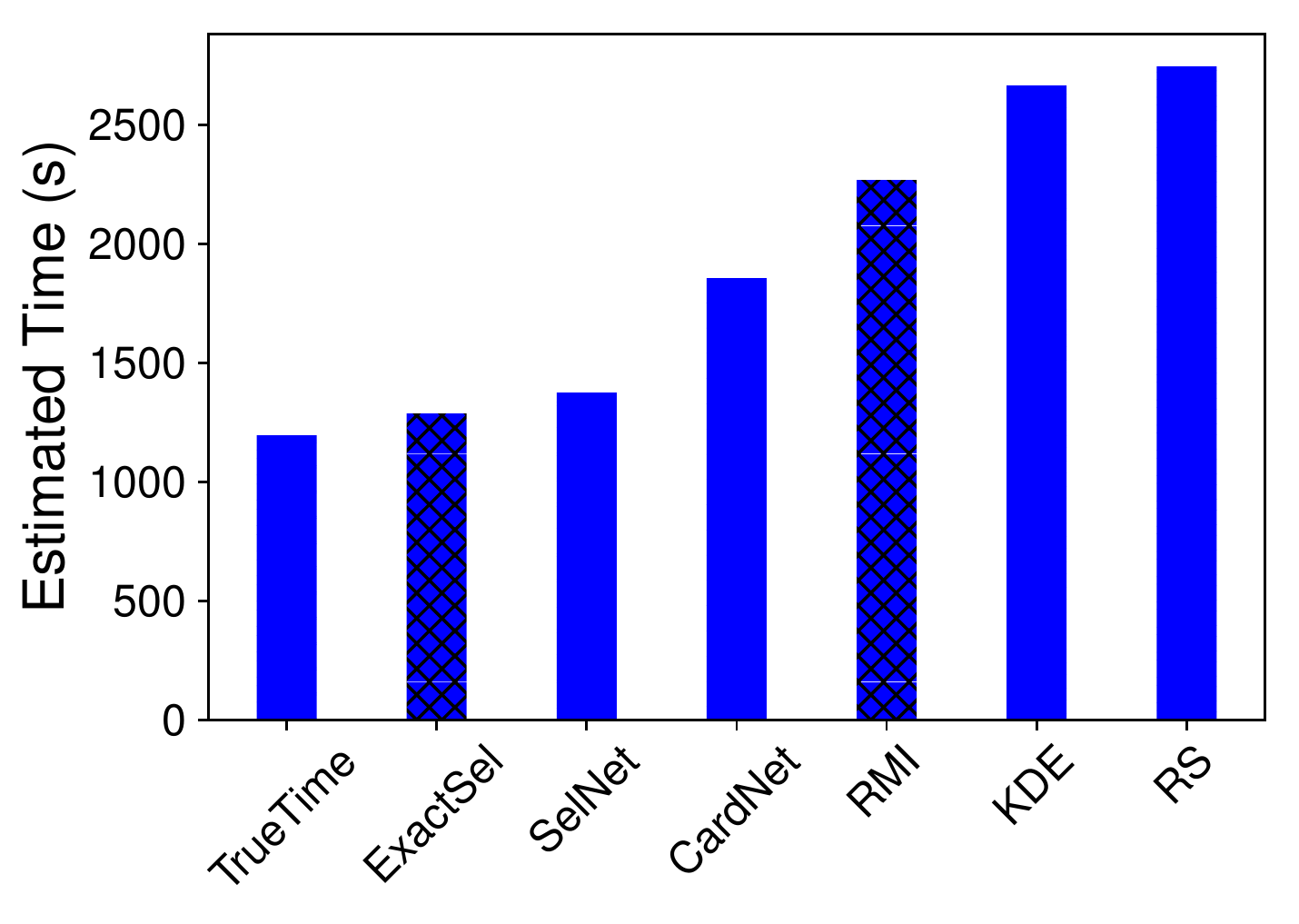}
    \label{fig:exp-application-2}
  }  
  \caption{Estimated search time (10,000 queries).}
  \vspace{-4ex}
  \label{fig:exp-application}
\end{figure}

To evaluate the usefulness of \selnetrp, we consider estimating the overall 
processing time for a query workload of semantic search: given a query text 
entry, we want to find matching records in the database. Estimating the query 
processing time may help to create a service level agreement. We use two 
datasets, \aminer publications (2.1M records) and \quora questions (0.8M 
records). \aminer has four attributes: title, authors, venue, and year. We 
follow \cite{DBLP:journals/pvldb/LiLSDT20} and concatenate attribute names 
and values as one string. \quora has one attribute. Then we embed each 
record to a 768-dimensional vector by 
\textsf{Sentence-BERT}~\cite{DBLP:conf/emnlp/ReimersG19}. 

10,000 records are sampled from each dataset as queries. To process a query, we 
first embed it by \textsf{Sentence-BERT}~\cite{DBLP:conf/emnlp/ReimersG19}, and 
then use \faiss~\cite{faiss} to find candidate records whose cosine similarity 
to the query embedding is no less than 0.9. The candidates are verified using 
\textsf{DITTO}~\cite{DBLP:journals/pvldb/LiLSDT20}. Hence the overall query 
processing time can be estimated as: $\text{avg\_\textsf{Faiss}\_time} \times 10000 
+ \text{avg\_\textsf{DITTO}\_time} \times \text{\faiss\_recall } \times
\sum_1^{10000} \text{estimated\_selectivity\_of\_query\_}i$. The average times and 
\faiss recall are obtained by running a small query workload. For selectivity, 
we consider \rs, \kde, \rmi, \cardnet, \selnetrp, and an oracle that outputs the 
exact selectivity (\exactsel). 

We plot the estimated time of processing 10,000 queries in Figure~\ref{fig:exp-application}, 
where \truetime indicates the ground truth. The models tend to underestimate on 
\aminer and overestimate on \quora. \selnetrp{}'s high accuracy in selectivity 
estimation pays off. Compared to the ground truth, \selnetrp{}'s error is 13\% on 
\aminer and 16\% on \quora, close to \exactsel's and much lower than the other 
models' (at least 39\%). \selnetrp is also efficient; e.g., running the workload 
to obtain the ground truth on \aminer spends 13 hours, which is twice the time of 
preparing training data $+$ training \selnetrp $+$ estimating for 10,000 queries. 
Seeing \selnetrp's scalability in estimation time (Table~\ref{tab:time}) and 
training time (Table~\ref{tab:traintime}), we believe that the advantage will be 
more substantial on larger datasets. 
}

%%!! DO NOT DELETE the following block

%%% Local Variables:
%%% mode: latex
%%% TeX-master: "paper"
%%% End:

\section{Conclusion}
\label{sec:con}
We tackled the selectivity estimation problem for high-dimensional data. 
Our method is based on learning monotonic query-dependent 
piece-wise linear function. This provides the flexibility of our model to approximate 
the selectivity curve while guaranteeing the consistency of estimation. We proposed a 
partitioning technique to cope with large-scale datasets and an incremental learning 
technique for updates. Our experiments showed the superiority of the proposed model in 
accuracy across a variety of datasets, distance functions, and error 
metrics. The experiments also demonstrated the usefulness of our model %benefits the query processing time estimation 
in a semantic search application.

%% DL to handle high-dim?

% In conclusion, we design a novel network architecture based on
% piece-wise linear function. First, we fully
% learns a good set of control points in threshold, and designs a
% function to learn the incremental interpolation predictions. Adopting a
% powerful function---neural networks in this work---to learn the
% interpolations results in enough flexibility for accurate
% estimation. Furthermore, our data-dependent control point model boosts
% the accuracy, and bring further flexibility. Due to our delicately
% designed model structure, we achieve the consistency, such that the
% increasing thresholds lead to the non-decreasing predictions. Finally,
% the partition-based estimator further improves the accuracy by
% alleviating the difficulty of each local model. The experimental
% results indicate the significant improvements of our model against the
% state-of-the-art approaches. 

%%% Local Variables:
%%% mode: latex
%%% TeX-master: "paper"
%%% End:

\noindent\textbf{Acknowledgements}\hspace{0.5em}%
This work was supported by NSFC 62072311 and U2001212, Guangdong 
Basic and Applied Basic Research Foundation 2019A1515111047 and 
2020B1515120028, Guangdong Peral River Recruitment Program of 
Talents 2019ZT08X603, JSPS Kakenhi 16H01722, 17H06099, 18H04093, 
and 19K11979, and ARC DPs 170103710 and 180103411. 

\balance

\bibliographystyle{abbrv}
\bibliography{citation}

\begin{thebibliography}{10}

\bibitem{URL:fasttext}
\url{https://fasttext.cc/docs/en/english-vectors.html}.

\bibitem{URL:glove}
\url{https://nlp.stanford.edu/projects/glove/}.

\bibitem{URL:youtube-faces}
\url{http://www.cs.tau.ac.il/~wolf/ytfaces/index.html}.

\bibitem{URL:deep}
\url{http://http://sites.skoltech.ru/compvision/noimi/}.

\bibitem{URL:sift}
\url{http://http://corpus-texmex.irisa.fr/}.

\bibitem{source-code}
\url{https://github.com/yyssl88/SelNet-Estimation}.

\bibitem{DBLP:conf/icdm/Anagnostopoulos15a}
C.~Anagnostopoulos and P.~Triantafillou.
\newblock Learning set cardinality in distance nearest neighbours.
\newblock In {\em {ICDM}}, pages 691--696, 2015.

\bibitem{DBLP:journals/tkdd/Anagnostopoulos17}
C.~Anagnostopoulos and P.~Triantafillou.
\newblock Query-driven learning for predictive analytics of data subspace
  cardinality.
\newblock {\em {ACM} Trans. Knowl. Discov. Data}, 11(4):47:1--47:46, 2017.

\bibitem{DBLP:conf/soda/AryaMM05}
S.~Arya, T.~Malamatos, and D.~M. Mount.
\newblock Space-time tradeoffs for approximate spherical range counting.
\newblock In {\em {SODA}}, pages 535--544, 2005.

\bibitem{DBLP:journals/is/AumullerBF20}
M.~Aum{\"{u}}ller, E.~Bernhardsson, and A.~J. Faithfull.
\newblock Ann-benchmarks: {A} benchmarking tool for approximate nearest
  neighbor algorithms.
\newblock {\em Inf. Syst.}, 87, 2020.

\bibitem{DBLP:conf/sigmod/BreunigKNS00}
M.~M. Breunig, H.~Kriegel, R.~T. Ng, and J.~Sander.
\newblock {LOF:} identifying density-based local outliers.
\newblock In {\em SIGMOD}, pages 93--104, 2000.

\bibitem{charikar2002similarity}
M.~S. Charikar.
\newblock Similarity estimation techniques from rounding algorithms.
\newblock In {\em STOC}, pages 380--388, 2002.

\bibitem{chen2016xgboost}
T.~Chen and C.~Guestrin.
\newblock Xgboost: A scalable tree boosting system.
\newblock In {\em KDD}, pages 785--794, 2016.

\bibitem{DBLP:journals/ftdb/CormodeGHJ12}
G.~Cormode, M.~N. Garofalakis, P.~J. Haas, and C.~Jermaine.
\newblock Synopses for massive data: Samples, histograms, wavelets, sketches.
\newblock {\em Foundations and Trends in Databases}, 4(1-3):1--294, 2012.

\bibitem{daniels2010monotone}
H.~Daniels and M.~Velikova.
\newblock Monotone and partially monotone neural networks.
\newblock {\em IEEE Transactions on Neural Networks}, 21(6):906--917, 2010.

\bibitem{DBLP:conf/sigmod/DasCDNKDARP17}
S.~Das, P.~S.~G. C., A.~Doan, J.~F. Naughton, G.~Krishnan, R.~Deep, E.~Arcaute,
  V.~Raghavendra, and Y.~Park.
\newblock Falcon: Scaling up hands-off crowdsourced entity matching to build
  cloud services.
\newblock In {\em {SIGMOD}}, pages 1431--1446, 2017.

\bibitem{fard2016fast}
M.~M. Fard, K.~Canini, A.~Cotter, J.~Pfeifer, and M.~Gupta.
\newblock Fast and flexible monotonic functions with ensembles of lattices.
\newblock In {\em NIPS}, pages 2919--2927, 2016.

\bibitem{Fox02robustregression}
J.~Fox.
\newblock Robust regression: Appendix to an r and s-plus companion to applied
  regression, 2002.

\bibitem{garcia2009lattice}
E.~Garcia and M.~Gupta.
\newblock Lattice regression.
\newblock In {\em NIPS}, pages 594--602, 2009.

\bibitem{guo2016msceleb}
Y.~Guo, L.~Zhang, Y.~Hu, X.~He, and J.~Gao.
\newblock M{S}-{C}eleb-1{M}: A dataset and benchmark for large scale face
  recognition.
\newblock In {\em ECCV}, 2016.

\bibitem{gupta2016monotonic}
M.~Gupta, A.~Cotter, J.~Pfeifer, K.~Voevodski, K.~Canini, A.~Mangylov,
  W.~Moczydlowski, and A.~Van~Esbroeck.
\newblock Monotonic calibrated interpolated look-up tables.
\newblock {\em The Journal of Machine Learning Research}, 17(1):3790--3836,
  2016.

\bibitem{han2017isotonic}
Q.~Han, T.~Wang, S.~Chatterjee, and R.~J. Samworth.
\newblock Isotonic regression in general dimensions.
\newblock {\em arXiv preprint arXiv:1708.09468}, 2017.

\bibitem{DBLP:conf/sigmod/HasanTAK020}
S.~Hasan, S.~Thirumuruganathan, J.~Augustine, N.~Koudas, and G.~Das.
\newblock Deep learning models for selectivity estimation of multi-attribute
  queries.
\newblock In {\em {SIGMOD}}, pages 1035--1050, 2020.

\bibitem{DBLP:conf/sigmod/HeimelKM15}
M.~Heimel, M.~Kiefer, and V.~Markl.
\newblock Self-tuning, {GPU}-accelerated kernel density models for
  multidimensional selectivity estimation.
\newblock In {\em {SIGMOD}}, pages 1477--1492, 2015.

\bibitem{huber1964robust}
P.~J. Huber et~al.
\newblock Robust estimation of a location parameter.
\newblock {\em The annals of mathematical statistics}, 35(1):73--101, 1964.

\bibitem{ioannidis2003history}
Y.~Ioannidis.
\newblock The history of histograms (abridged).
\newblock In {\em VLDB}, pages 19--30, 2003.

\bibitem{fastercovertree}
M.~Izbicki and C.~R. Shelton.
\newblock Faster cover trees.
\newblock In {\em {ICML}}, pages 1162--1170, 2015.

\bibitem{faiss}
H.~J{\'{e}}gou, tthijs Douze, and J.~Johnson.
\newblock Facebook ai similarity search (faiss).
\newblock \url{https://github.com/facebookresearch/faiss}.

\bibitem{DBLP:conf/cidr/KipfKRLBK19}
A.~Kipf, T.~Kipf, B.~Radke, V.~Leis, P.~A. Boncz, and A.~Kemper.
\newblock Learned cardinalities: Estimating correlated joins with deep
  learning.
\newblock In {\em {CIDR}}, 2019.

\bibitem{DBLP:conf/sigmod/KipfMRSKK020}
A.~Kipf, R.~Marcus, A.~van Renen, M.~Stoian, A.~Kemper, T.~Kraska, and
  T.~Neumann.
\newblock Radixspline: a single-pass learned index.
\newblock In {\em aiDM@SIGMOD}, pages 5:1--5:5, 2020.

\bibitem{kraska2018case}
T.~Kraska, A.~Beutel, E.~H. Chi, J.~Dean, and N.~Polyzotis.
\newblock The case for learned index structures.
\newblock In {\em SIGMOD}, pages 489--504, 2018.

\bibitem{lathuiliere2018comprehensive}
S.~Lathuili{\`e}re, P.~Mesejo, X.~Alameda-Pineda, and R.~Horaud.
\newblock A comprehensive analysis of deep regression.
\newblock {\em arXiv preprint arXiv:1803.08450}, 2018.

\bibitem{DBLP:conf/sigmod/Li0ZY020}
P.~Li, H.~Lu, Q.~Zheng, L.~Yang, and G.~Pan.
\newblock {LISA:} {A} learned index structure for spatial data.
\newblock In {\em {SIGMOD}}, pages 2119--2133, 2020.

\bibitem{DBLP:journals/tkde/LiZSWLZL20}
W.~Li, Y.~Zhang, Y.~Sun, W.~Wang, M.~Li, W.~Zhang, and X.~Lin.
\newblock Approximate nearest neighbor search on high dimensional data -
  experiments, analyses, and improvement.
\newblock {\em {IEEE} Trans. Knowl. Data Eng.}, 32(8):1475--1488, 2020.

\bibitem{DBLP:journals/pvldb/LiLSDT20}
Y.~Li, J.~Li, Y.~Suhara, A.~Doan, and W.-C. Tan.
\newblock Deep entity matching with pre-trained language models.
\newblock {\em {PVLDB}}, 14(1):50--60, 2020.

\bibitem{DBLP:conf/edbt/MattigFBS18}
M.~Mattig, T.~Fober, C.~Beilschmidt, and B.~Seeger.
\newblock Kernel-based cardinality estimation on metric data.
\newblock In {\em EDBT}, pages 349--360, 2018.

\bibitem{novelinkova2011comparison}
M.~Novelinkova.
\newblock Comparison of clenshaw-curtis and gauss quadrature.
\newblock In {\em WDS}, volume~11, pages 67--71, 2011.

\bibitem{DBLP:journals/corr/abs-1905-06425}
J.~Ortiz, M.~Balazinska, J.~Gehrke, and S.~S. Keerthi.
\newblock An empirical analysis of deep learning for cardinality estimation.
\newblock {\em CoRR}, abs/1905.06425, 2019.

\bibitem{DBLP:conf/sigmod/ParkZM20}
Y.~Park, S.~Zhong, and B.~Mozafari.
\newblock Quicksel: Quick selectivity learning with mixture models.
\newblock In {\em {SIGMOD}}, pages 1017--1033, 2020.

\bibitem{prunty1983curve}
L.~Prunty.
\newblock Curve fitting with smooth functions that are piecewise-linear in the
  limit.
\newblock {\em Biometrics}, pages 857--866, 1983.

\bibitem{DBLP:conf/emnlp/ReimersG19}
N.~Reimers and I.~Gurevych.
\newblock Sentence-{BERT}: Sentence embeddings using siamese {BERT}-networks.
\newblock In {\em {EMNLP-IJCNLP}}, pages 3980--3990, 2019.

\bibitem{schroff2015facenet}
F.~Schroff, D.~Kalenichenko, and J.~Philbin.
\newblock Facenet: A unified embedding for face recognition and clustering.
\newblock In {\em CVPR}, pages 815--823, 2015.

\bibitem{shazeer2017outrageously}
N.~Shazeer, A.~Mirhoseini, K.~Maziarz, A.~Davis, Q.~Le, G.~Hinton, and J.~Dean.
\newblock Outrageously large neural networks: The sparsely-gated
  mixture-of-experts layer.
\newblock {\em arXiv preprint arXiv:1701.06538}, 2017.

\bibitem{spouge2003least}
J.~Spouge, H.~Wan, and W.~Wilbur.
\newblock Least squares isotonic regression in two dimensions.
\newblock {\em Journal of Optimization Theory and Applications},
  117(3):585--605, 2003.

\bibitem{DBLP:journals/pvldb/SunL19}
J.~Sun and G.~Li.
\newblock An end-to-end learning-based cost estimator.
\newblock {\em {PVLDB}}, 13(3):307--319, 2019.

\bibitem{sun2013deep}
Y.~Sun, X.~Wang, and X.~Tang.
\newblock Deep convolutional network cascade for facial point detection.
\newblock In {\em CVPR}, pages 3476--3483, 2013.

\bibitem{toshev2014deeppose}
A.~Toshev and C.~Szegedy.
\newblock Deeppose: Human pose estimation via deep neural networks.
\newblock In {\em CVPR}, pages 1653--1660, 2014.

\bibitem{DBLP:journals/pvldb/WalenzSRY19}
B.~Walenz, S.~Sintos, S.~Roy, and J.~Yang.
\newblock Learning to sample: Counting with complex queries.
\newblock {\em {PVLDB}}, 13(3):390--402, 2019.

\bibitem{DBLP:conf/icbsp/WangZZ17}
D.~Wang, Y.~Zhang, and Y.~Zhao.
\newblock Lightgbm: An effective mirna classification method in breast cancer
  patients.
\newblock In {\em ICCBB}, pages 7--11, 2017.

\bibitem{DBLP:conf/sigmod/WangXQ0SWO20}
Y.~Wang, C.~Xiao, J.~Qin, X.~Cao, Y.~Sun, W.~Wang, and M.~Onizuka.
\newblock Monotonic cardinality estimation of similarity selection: {A} deep
  learning approach.
\newblock In {\em {SIGMOD}}, pages 1197--1212, 2020.

\bibitem{unconstraintmono}
A.~Wehenkel and G.~Louppe.
\newblock Unconstrained monotonic neural networks.
\newblock In {\em NeurIPS}, pages 1543--1553, 2019.

\bibitem{whang1994dynamic}
K.-Y. Whang, S.-W. Kim, and G.~Wiederhold.
\newblock Dynamic maintenance of data distribution for selectivity estimation.
\newblock {\em {VLDB} J.}, 3(1):29--51, 1994.

\bibitem{wu2016sampling}
W.~Wu, J.~F. Naughton, and H.~Singh.
\newblock Sampling-based query re-optimization.
\newblock In {\em {SIGMOD}}, pages 1721--1736, 2016.

\bibitem{wu2018local}
X.~Wu, M.~Charikar, and V.~Natchu.
\newblock Local density estimation in high dimensions.
\newblock In {\em ICML}, pages 5293--5301, 2018.

\bibitem{DBLP:conf/icde/WuAA02}
Y.~Wu, D.~Agrawal, and A.~{El Abbadi}.
\newblock Query estimation by adaptive sampling.
\newblock In {\em ICDE}, pages 639--648, 2002.

\bibitem{DBLP:journals/pvldb/YangLKWDCAHKS19}
Z.~Yang, E.~Liang, A.~Kamsetty, C.~Wu, Y.~Duan, P.~Chen, P.~Abbeel, J.~M.
  Hellerstein, S.~Krishnan, and I.~Stoica.
\newblock Deep unsupervised cardinality estimation.
\newblock {\em {PVLDB}}, 13(3):279--292, 2019.

\bibitem{you2017deep}
S.~You, D.~Ding, K.~Canini, J.~Pfeifer, and M.~Gupta.
\newblock Deep lattice networks and partial monotonic functions.
\newblock In {\em NIPS}, pages 2981--2989, 2017.

\bibitem{zhang2018word}
D.~Zhang and Z.~Yang.
\newblock Word embedding perturbation for sentence classification.
\newblock {\em arXiv preprint arXiv:1804.08166}, 2018.

\end{thebibliography}

\fullversion{
\newpage
\appendix
\section*{Appendix}

\section{Proof} \label{sec:proof}

\paragraph{Lemma~\ref{lm:mono}}
\begin{proof}
Assume $t \in [\tau_{i-1}, \tau_i)$, then $t + \epsilon$ is 
in $[\tau_{i-1}, \tau_i)$ or $[\tau_i, \tau_{i+1})$.
In the first case, $\est{f}(\xx, t + \epsilon, \mathcal{D}; \Theta) - \est{f}(\xx, t, \mathcal{D}; \Theta) = \frac{\epsilon}{\tau_i - \tau_{i-1}} \cdot (p_i - p_{i-1}) \geq 0$. In the second case, 
$\est{f}(\xx, t, \mathcal{D}; \Theta) \leq p_i$ and $\est{f}(\xx, t + \epsilon, \mathcal{D}; \Theta) \geq p_i$. 
Therefore, $\est{f}(\xx, t, \mathcal{D}; \Theta)$ is non-decreasing in $t$.
\end{proof}

% \paragraph{Lemma~\ref{lm:flex}.}
% \begin{proof}
% We discuss the approximation power of a piecewise linear function. 
% Given a database $\mathcal{D}$ and a query object $\xx$, for any threshold $t$, we have 
% $|f(\xx, t, \mathcal{D}) - \est{f}(\xx, t, \mathcal{D}; \Theta)| = K \cdot |\Delta| \cdot |\nabla \est{f}|$, 
% where $K$ is a constant value and $|\Delta| = \max_{i=0}^L\{\tau_{i+1} - \tau_i\}$. 
% Because $\est{f}(\xx, t, \mathcal{D}; \Theta)$ is a piecewise linear function, $|\nabla \est{f}|$ is bounded
% by $\max_{i=0}^L\{\frac{p_{i+1} - p_i}{\tau_{i+1} - \tau_i}\}$. With sufficiently large 
% number of control points and optimal $\{\tau_i\}_{i=1}^L$ and $\{p_i\}_{i=0}^{L+1}$, 
% both $|\Delta|$ and $|\nabla \est{f}|$ can be arbitrarily small. 
% Therefore, for any $\epsilon > 0$, $K \cdot |\Delta| \cdot |\nabla \est{f}| < \epsilon$. 
% % Second, $\{\tau_i\}_{i=1}^L$
% % and $\{p_i\}_{i=0}^{L+1}$ are estimated by neural networks, and $K \cdot |\Delta||\nabla \est{f}|$ 
% % is bounded by $K \cdot (|\Delta| + \epsilon_1 )\cdot (|\nabla \est{f}| + \epsilon_2)$. Due to the universal
% % approximation of neural networks, there exist $\epsilon_1$ and $\epsilon_2$ such that
% % $K \cdot (|\Delta| + \epsilon_1) (|\nabla \est{f}| + \epsilon_2) < \epsilon$.
% \end{proof}

\section{Experiment Setup} \label{sec:exp-setup}

% \subsection{Data Preparation}%
% % \fixmelater{Only few $y$ values. Regression becomes classification}
% %% !! , remove duplicates???
% We randomly sample 0.25 million unique vectors as queries from the dataset $\mathcal{D}$. %
% %%!! mention that thresholds needs to be different for different x
% In order to generate more representative thresholds for each vector, we follow
% the approach in~\cite{DBLP:conf/edbt/MattigFBS18}: we generate a geometric
% sequence of $w$ selectivity values in the range
% $[1, \frac{|\mathcal{D}|}{100}]$, and calculate their corresponding thresholds.
% The resulting data are randomly split 80:10:10 \emph{according to queries}
% into training (\train), validation (\valid) and test data (\test). This ensures that none of the test
% queries and their thresholds has been seen by the model during the training or
% validation. We use $w = 40$. 

\subsection{Model Settings} \label{sec:hyper}
%We provide the model settings in the experiments. 
Hyperparameter and training settings are given below. 
\begin{itemize}
  \item \is and \kde: The sample size is 2000. 
  \item \qra and \qrb: The number of query prototypes is 2000.
  \item \lightgbm: The number of CARTs is 1000.
  \item \dnn is a vanilla FFN with four
  hidden layers of sizes 512, 512, 512, and 256.
  \item \moe consists of 30 expert models, each an FFN with three hidden
  layers of sizes 512, 512, and 512. We used top-3 experts for the prediction. 
  \item \rmi has three levels, with 1, 4, and 8 models, respectively. Each 
  model is an FFN with four hidden layers with sizes 512, 512, 512, and 256.
  \item \dln is an architecture of six layers: calibrators, linear embedding, 
  calibrators, ensemble of lattices, calibrators, and linear embedding. 
  \item \umnn is an FFN with four hidden layers of sizes 512, 512, 512 and 256 to 
  implement the derivative. 
  $\frac{\partial f(\xx, t, \mathcal{D})}{\partial t}$. $f(\xx, t, \mathcal{D})$ 
  is computed by Clenshaw-Curtis quadrature with learned derivatives.
  \item \selnetrp: We use an FFN with two hidden layers to estimate $\Btau$, and 
  an FFN in Equation~\ref{eq:modelm} with four hidden layers to estimate $\pp$. 
  The encoder and decoder of AE are implemented with an FFN with three hidden layers.
  For \face{} and \youtube{}, the sizes of the first three (or two, if it only has two) 
  hidden layers of the three FFNs are 512, and the sizes of all the other hidden layers 
  are 256. For \fasttext{}, \glove{}, \deep{}, and \sift{}, the sizes of the first hidden 
  layer of the these FFNs are 1024, and the others remain the same as above. 
  The number of control parameters $L$ is 50. The default partition size $K$ is 3. 
  $t_{\max}$ is 54 for Euclidean distance. For cosine similarity, we equivalently convert 
  it to Euclidean distance on unit vectors, and set $t_{\max} = 1$. 
  The learning rates of \face{}, \fasttext{}, \youtube{}, \glove{}, \deep{}, and \sift{} are
  0.00003, 0.00002, 0.00003, 0.0001, 0.0001, and 0.0001, respectively. 
  $|\mathbf{h}_i|$ ($0 \leq i \leq L + 1$) in model $M$ is 100. 
  The batch size is 512 for all the datasets. We train all the models in 1500 epochs and 
  select the ones with the smallest validation error. 
  For training with data partitioning, we use $T = 300$ and $\beta = 0.1$. 
  $\delta_U$ for incremental learning is 20.
\end{itemize}

% For \is and \kde, we use 2,000 samples to keep the estimation cost
% reasonable. 
%
%\kde only supports metric distance functions. 
%but Cosine distance violates the property of
%triangle inequality.
%
For the learning models, we train them with the same Huber loss over the logarithms 
of the ground truth and the predicted value. All the hyper-parameters are fine-tuned 
to minimize the validation error. %
\dnn{}, \moe{} and \rmi{} cannot directly handle the threshold $t$. We learn a
non-linear transformation of $t$ into an $m$-dimensional embedding vector, i.e.,
$\tt = \relu(\ww t)$. %
Then we concatenate it with $\xx$ as the input to these models. 

\subsection{Evaluation Metrics} \label{sec:metric}
We evaluate Mean Squared Error (\mse), Mean Absolute Error (\mae), and Mean
Absolute Percentage Error (\mre). They are defined as:
\begin{align*}
  & \mathsf{MSE} = \frac{1}{m}\sum_{i=1}^m (\est{y}_i - y_i)^2, \\
  & \mathsf{MAE} = \frac{1}{m}\sum_{i=1}^m \abs{\est{y}_i - y_i}, \\
  %& \mathsf{MAPE}  = \frac{1}{m}\sum_{i=1}^m \abs{\frac{\est{y}_i - y_i}{y_i + \varepsilon}}
  & \mathsf{MAPE}  = \frac{1}{m}\sum_{i=1}^m \abs{\frac{\est{y}_i - y_i}{y_i}}, 
\end{align*}
where $y_i$ is the ground truth value and $\est{y}_i$ is the estimated value.

% \subsection{Environment}
% The experiments were carried out on a server with a Intel Xeon E5-2640 @2.40GHz 
% CPU and 256GB RAM running Ubuntu 16.04.4 LTS. Non-deep models were implemented 
% in C++. Deep models were implemented in python and Tensorflow. 

%Source codes and datasets are available at \url{https://github.com/yaoshuwang/SelNet-Estimation}.
}

\end{document}